\documentclass[11pt]{article}
\usepackage{amsmath,amsthm,amscd,array,amssymb}

\begin{document}

\begin{center}
{\Large{\bf  $N_T = 4$ equivariant extension of the 
\\
\medskip\smallskip
$3D$ topological model of Blau and Thompson}}
\\
\bigskip\medskip
{\large{\sc B. Geyer}}$^a$
\footnote{Email: geyer@itp.uni-leipzig.de}
and 
{\large{\sc D. M\"ulsch}}$^{b}$
\footnote{Email: muelsch@informatik.uni-leipzig.de}
\\
\smallskip
{\it $^a$ Universit\"at Leipzig, Naturwissenschaftlich-Theoretisches Zentrum
\\
$~$ and Institut f\"ur Theoretische Physik, D--04109 Leipzig, Germany
\\\smallskip
$\!\!\!\!\!^b$ Wissenschaftszentrum Leipzig e.V., D--04103 Leipzig, Germany}
\\
\bigskip
{\small{\bf Abstract}}
\\
\end{center}

\begin{quotation}
\noindent {\small{
The Blau--Thompson $N_T = 2$, $D = 3$ non--equivariant topological model, 
obtained through the so--called ,novel' twist of $N = 4$, $D = 3$ 
super Yang--Mills theory, is extended to a $N_T = 4$, $D = 3$ topological 
theory. The latter, formally, may be regarded as a topological non--trivial 
deformation of the $N_T = 2$, $D = 4$ Yamron--Vafa--Witten theory after 
dimensional reduction to $D = 3$. For completeness also the dimensional
reduction of the half--twisted $N_T = 2$, $D = 4$ Yamron model is explicitly
constructed.
}}
\end{quotation}

\medskip
\begin{flushleft}
{\large{\bf 1. Introduction}}
\end{flushleft}
\medskip
Topological quantum field theory (TQFT) has become an interesting link between 
physics and mathematics. It has connected diverse areas and many of the  
advanced ideas in QFT and string theory with the ones involved in topology
(see, e.g., Refs.~\cite{1} -- \cite{13}). TQFTs with simple, $N_T = 1$, 
topological symmetry have been widely studied in different space--time 
dimensions, e.g., the topological sigma models in $D = 2$ \cite{14}, the 
Chern--Simons gauge theory in $D = 3$ \cite{4} and the Donaldson--Witten 
theory in $D = 4$ \cite{3}, namely, from both the perturbative and the 
non--perturbative point of view. TQFTs with extended, $N_T > 1$, 
topological symmetry have also been considered, e.g., as effective 
world volume theories of D3--branes \cite{15}, D2--branes \cite{16} and 
M5--branes \cite{17} in string theory wrapping supersymmetric cycles 
of higher dimensional compactification manifolds. They provide a promising 
arena for testing key ideas as $S$--duality \cite{18,19}, large $N$--dynamics 
of supersymmetric gauge theories and, eventually, 
the AdS/CFT conjecture \cite{20}. 

Usually, the topological supersymmetry is realized equivariantly, i.e.,
prior to the introduction of gauge ghosts the cohomology closes only modulo
equations of motions. However, by introducing the ,novel' topological twist
of the $N = 4$, $D = 3$ super Yang--Mills theory (SYM) Blau and Thompson
\cite{16} obtained a $N_T = 2$, $D = 3$ topological model whose topological 
shift 
symmetry is strictly nilpotent even prior to the introduction of the gauge 
ghosts. Such theories are intrinsic for $D = 3$ and, obviously, quite special. 
After carrying out a dimensional reduction to $D = 2$ one obtains a $N_T = 4$ 
Hodge--type cohomological theory \cite{21}. In such theories, completely 
analogous to de Rham cohomology, there exists besides the topological shift 
operator also a co--shift operator both of which are nilpotent and
interrelated by a discrete Hodge--type duality. 

Motivated by this interesting possibility we looked for an equivariant
$N_T = 4$ extension of the Blau--Thompson model which might provide, after
dimensional reduction, another $N_T = 8$ Hodge--type cohomological theory. 
If such an equivariant extension would exist it had to be equivalent to
one of the various non-equivalent $N_T \geq 2$, $D = 3$ topological theories.

\pagebreak


\setlength{\unitlength}{1.mm}

\begin{picture}(160,230)
\thicklines 
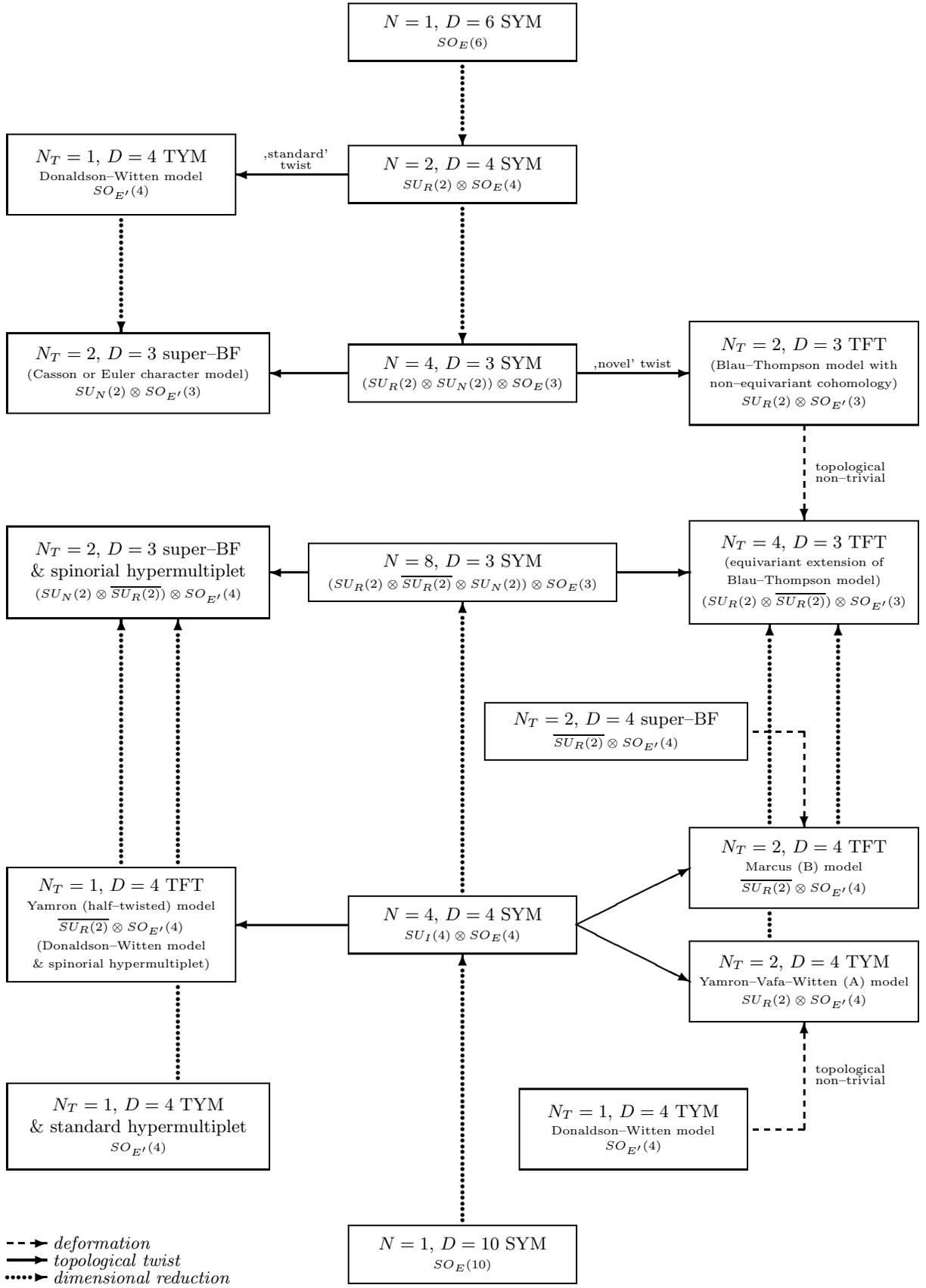
\put(0,3){\footnotesize {\sc Diagram 1:} Different twisting of $N = 2$, 
$N = 4$ and $N = 8$ SYM theories and}
\put(0,0){\footnotesize \phantom{\sc Diagram 1:} interrelations of 
various topological theories in $D = 3$ and $D = 4$.}
\thicklines
\multiput(0,16)(1,0){5}{\circle*{0.6}}
\put(5,16){\vector(1,0){2}}
\put(8,15){\footnotesize {\it dimensional reduction}}
\thicklines
\put(0,19){\vector(1,0){7}}
\put(8,18){\footnotesize {\it topological twist}}
\thicklines
\multiput(0,22)(2,0){3}{\line(1,0){1}}
\put(5,22){\vector(1,0){2}}
\put(8,21){\footnotesize {\it deformation}}
\thinlines
\put(0,203){\framebox(40,14){\shortstack{\footnotesize
            $N_T = 1$, $D = 4$ TYM \\ \tiny
            Donaldson--Witten model \\ \tiny
            $SO_{E^\prime}(4)$}}}
\thicklines
\multiput(20,202)(0,-1){18}{\circle*{0.6}}
\put(20,184){\vector(0,-1){2}}
\thinlines
\put(0,168){\framebox(46,14){\shortstack{\footnotesize
            $N_T = 2$, $D = 3$ super--BF \\ \tiny
            (Casson or Euler character model) \\ \tiny
            $SU_N(2) \otimes SO_{E^\prime}(3)$}}}
\put(0,132){\framebox(46,16){\shortstack{\footnotesize
            $N_T = 2$, $D = 3$ super--BF \\ \footnotesize
            $\&$ spinorial hypermultiplet \\ \tiny
            $(SU_N(2) \otimes \overline{SU_R(2)}) \otimes SO_{E^\prime}(4)$}}}
\put(60,230){\framebox(40,10){\shortstack{\footnotesize
             $N = 1$, $D = 6$ SYM \\ \tiny
             $SO_E(6)$}}}
\thicklines
\multiput(80,229)(0,-1){12}{\circle*{0.6}}
\put(80,217){\vector(0,-1){2}}
\thinlines
\put(60,205){\framebox(40,10){\shortstack{\footnotesize
             $N = 2$, $D = 4$ SYM \\ \tiny
             $SU_R(2) \otimes SO_E(4)$}}}
\thicklines
\put(45,213){\tiny ,standard'}
\put(48,211){\tiny twist}
\multiput(80,204)(0,-1){22}{\circle*{0.6}}
\put(80,182){\vector(0,-1){2}}
\thinlines
\thicklines
\put(60,210){\vector(-1,0){20}}
\thinlines
\put(60,170){\framebox(40,10){\shortstack{\footnotesize 
             $N = 4$, $D = 3$ SYM \\ \tiny
             $(SU_R(2) \otimes SU_N(2)) \otimes SO_E(3)$}}}
\put(103,176){\tiny ,novel' twist}
\thicklines
\put(60,175){\vector(-1,0){14}}
\put(100,175){\vector(1,0){20}}
\thinlines
\put(53,135){\framebox(54,10){\shortstack{\footnotesize 
             $N = 8$, $D = 3$ SYM \\ \tiny
             $(SU_R(2) \otimes \overline{SU_R(2)} 
             \otimes SU_N(2)) \otimes SO_E(3)$}}}
\thicklines
\put(53,140){\vector(-1,0){7}}
\put(107,140){\vector(1,0){13}}
\thinlines
\put(120,166){\framebox(40,18){\shortstack{\footnotesize 
              $N_T = 2$, $D = 3$ TFT \\ \tiny
              (Blau--Thompson model with \\ \tiny
              non--equivariant cohomology) \\ \tiny
              $SU_R(2) \otimes SO_{E^\prime}(3)$}}}
\put(142,158){\tiny topological} 
\put(142,156){\tiny non--trivial}
\thicklines
\multiput(140,151)(0,2){8}{\line(0,1){1}}
\put(140,151){\vector(0,-1){2}}
\thinlines
\put(120,131){\framebox(40,18){\shortstack{\footnotesize 
              $N_T = 4$, $D = 3$ TFT \\ \tiny
              (equivariant extension of \\ \tiny             
              Blau--Thompson model) \\ \tiny
              $(SU_R(2)\otimes \overline{SU_R(2)})\otimes SO_{E^\prime}(3)$}}}
\thicklines
\put(20,130){\vector(0,1){2}}
\multiput(20,89)(0,1){41}{\circle*{0.6}}
\thinlines
\put(0,68){\framebox(40,20){\shortstack{\footnotesize
           $N_T = 1$, $D = 4$ TFT \\ \tiny
           Yamron (half--twisted) model \\ \tiny
           $\overline{SU_R(2)} \otimes SO_{E^\prime}(4)$ \\ \tiny
           (Donaldson--Witten model \\ \tiny
           $\&$ spinorial hypermultiplet)}}}
\thicklines
\put(30,130){\vector(0,1){2}}
\multiput(30,89)(0,1){41}{\circle*{0.6}}
\multiput(30,51)(0,1){17}{\circle*{0.6}}
\thinlines
\put(0,35){\framebox(46,15){\shortstack{\footnotesize
           $N_T = 1$, $D = 4$ TYM \\ \footnotesize
           $\&$ standard hypermultiplet \\ \tiny
           $SO_{E^\prime}(4)$}}}
\thicklines
\put(80,133){\vector(0,1){2}}
\multiput(80,84)(0,1){49}{\circle*{0.6}}
\thinlines
\put(60,73){\framebox(40,10){\shortstack{\footnotesize
            $N = 4$, $D = 4$ SYM \\ \tiny
            $SU_I(4) \otimes SO_E(4)$}}}
\thicklines
\put(100,78){\vector(2,1){20}}
\put(100,78){\vector(2,-1){20}}
\put(60,78){\vector(-2,0){20}}
\put(80,71){\vector(0,1){2}}
\multiput(80,26)(0,1){45}{\circle*{0.6}}
\thinlines
\put(60,15){\framebox(40,10){\shortstack{\footnotesize
            $N = 1$, $D = 10$ SYM \\ \tiny
            $SO_E(10)$}}}
\put(84,107){\framebox(46,10){\shortstack{\footnotesize
             $N_T = 2$, $D = 4$ super--BF \\ \tiny
             $\overline{SU_R(2)} \otimes SO_{E^\prime}(4)$}}}
\thicklines
\multiput(131,112)(2,0){5}{\line(1,0){1}}
\multiput(140,111)(0,-2){9}{\line(0,1){1}}
\put(140,97){\vector(0,-1){2}}
\thinlines
\thicklines
\put(134,129){\vector(0,1){2}}
\multiput(134,96)(0,1){33}{\circle*{0.6}}
\put(146,129){\vector(0,1){2}}
\multiput(146,96)(0,1){33}{\circle*{0.6}}
\thinlines
\put(120,81){\framebox(40,14){\shortstack{\footnotesize
             $N_T = 2$, $D = 4$ TFT \\ \tiny
             Marcus (B) model \\ \tiny
             $\overline{SU_R(2)} \otimes SO_{E^\prime}(4)$}}}
\thicklines
\multiput(134,76)(0,1){5}{\circle*{0.6}}
\thinlines
\put(120,61){\framebox(40,14){\shortstack{\footnotesize
             $N_T = 2$, $D = 4$ TYM \\ \tiny
             Yamron--Vafa--Witten (A) model \\ \tiny
             $SU_R(2) \otimes SO_{E^\prime}(4)$}}}
\thicklines
\multiput(131,42)(2,0){5}{\line(1,0){1}}
\put(140,59){\vector(0,1){2}}
\multiput(140,42)(0,2){9}{\line(0,1){1}}
\thinlines
\put(142,52){\tiny topological} 
\put(142,50){\tiny non--trivial}
\put(90,35){\framebox(40,14){\shortstack{\footnotesize
            $N_T = 1$, $D = 4$ TYM \\ \tiny
            Donaldson--Witten model \\ \tiny
            $SO_{E^\prime}(4)$}}}
\end{picture}

\pagebreak

A complete group theoretical classification of all the topological twists of 
$N = 4$ and $N = 8$ super--Yang--Mills theory (SYM) in $D = 3$ has been given 
in \cite{16}. According to that analysis in both cases there exist exactly 
two possible topological twists (see Diagram). 

In the case $N = 4$, $D = 3$ SYM the ,standard' $N_T = 2$ topological 
twist gives the super--BF model whereas the so--called ,novel' $N_T = 2$ 
topological twist by construction leads to a model, henceforth called 
Blau--Thompson (BT) model, which precisely enjoys the above mentioned 
property, i.e., which has no bosonic scalar fields and hence no underlying 
equivariant cohomology. 

Also in Ref.~\cite{16} it has been shown, without constructing the
corresponding models explicitly, that there exist only two (partial) 
topological twists of $N = 8$, $D = 3$ SYM, provided one excludes 
theories involving higher spin fields. These models having an underlying 
$N_T = 4$ and $N_T = 2$ topological symmetry describe world--volume theories
of D2--brane instantons wrapping supersymmetric three--cycles of
Calabi--Yau three--folds and $G_2$--holonomy Joyce manifolds. 

Moreover, in \cite{16} it has been shown that the $N_T = 4$, $D = 3$ 
model is just the dimensional reduction of either of the two 
$N_T = 2$, $D = 4$ models to $D = 3$, namely, the so--called
$A$--model constructed by Yamron \cite{22} and Vafa--Witten \cite{18}
and the $B$--model constructed by Marcus \cite{23}, whereas the 
$N_T = 2$, $D = 3$ model arises from the dimensional reduction of the 
,half--twisted' $N_T = 1$, $D = 4$ theory \cite{22}.

In this paper we construct explicitly both the $N_T = 2$ and the $N_T = 4$ 
model in $D=3$ dimensions. Thereby, we restrict ourselves to Euclidean 
space--time.
In seaching of cohomological Hodge--type theories this restriction is 
convenient since, after performing a dimensional reduction to $D = 2$, the 
topological co--shift symmetry is partially encoded in the vector 
supersymmetry. Under that restriction both theories can be 
characterized uniquely by imposing besides the topological shift symmetry 
$Q^a$ (and possibly $\bar{Q}^a$) also the vector supersymmetry $Q^a_\alpha$ 
(and possibly $\bar{Q}^a_\alpha$). Furthermore, the latter model
coincides with the $N_T = 4$ equivariant extention of the Blau--Thompson model 
which we consider at first instance. The equivalence of both models obtains
by deforming explicitly the Yamron--Vafa--Witten theory, i.e., the $A$--model, 
after dimensional reduction to $D = 3$. In addition, we remark that we have 
not been able to find an appropiate equivariant extension of the novel 
topological model preserving the number of topological supercharges $N_T = 2$.
On the other hand, due to the completeness of the classification \cite{16}, 
such an extension should really not be possible.

The paper is organized as follows. Following Ref.~\cite{16}, in Section 2 
we recall the structure of the two possible topological twists of 
$N =  4$, $D = 3$ SYM, leading to the $N_T = 2$, $D = 3$ super--BF model 
\cite{24} -- \cite{26} and the novel $N_T = 2$, $D = 3$ Blau--Thompson model. 
In Section 3 we construct a $N_T = 4$ equivariant extension of the 
Blau Thompson model. In Section 4 
it is shown that this extension coincides with the Yamron--Vafa--Witten 
theory after carrying out a dimensional reduction to $D = 3$, i.e., with the 
$N_T = 4$ topological twist of $N = 8$, $D = 3$ SYM. In Section 5 
we construct the $N_T = 2$ topological model of $N = 8$, $D = 3$ SYM which 
is the extension of the $N_T = 2$, $D = 3$ super--BF model by a spinorial 
hypermultiplet.
    
\bigskip\medskip
\begin{flushleft}
{\large{\bf 2. The two topological twists of $N = 4$, $D = 3$ SYM theory: \\
~~~~Super--BF and Blau--Thompson model}}
\end{flushleft}
\bigskip

In this section we briefly recall the two possible topological twists 
of $N = 4$ SYM in $D = 3$ dimensional Euclidean space--time which obtains 
by dimensional reduction of $N = 1$, $D = 6$ SYM, either directly or via 
$N = 2$, $D = 4$ SYM, to $D = 3$ (cf., upper half of the Diagram). 
As pointed out in Ref.~\cite{27}, this theory has a global 
$(SU(2)_R \otimes SU(2)_N) \otimes SU(2)_E$ symmetry, where the $SU(2)_R$ 
group primarily results from the symmetry of the fermions of $N = 1$ SYM 
in $D = 6$. 

After dimensional reduction the gauge multiplet of 
$N = 4$ SYM in $D = 3$ contains three scalar fields which transform in the 
vector representation under the group $SU(2)_N$, the internal Euclidean 
symmetry group arising from the decomposition 
$Spin(6) \rightarrow SU(2)_N \otimes SU(2)_E$. The symmetry group 
$SU(2)_E$ is the Euclidean rotation group in $D = 3$.      

There are only two essentially different possibilities to construct 
topological models with $N_T = 2$ scalar topological supercharges, arising 
from twisting $N = 4$ SYM in $D = 3$ \cite{16}. 
\medskip

\noindent
{\it (A) The super--BF model}
\\
The standard twist consists in replacing $SU(2)_E \otimes SU(2)_R$ through 
its diagonal subgroup. This leads to the universal gauge multiplet 
$\{ A_\alpha, \psi_\alpha^a, \phi^{ab}, \eta^a \},\, a = 1,2;
\, \alpha = 1,2,3$, of the $N_T = 2$, $D = 3$ super--BF model 
\cite{24} -- \cite{26}, which is built up from the gauge field $A_\alpha$, 
a $SU(2)_N$ doublet of Grassmann--odd topological ghost--antighost vector 
fields $\psi_\alpha^a = \{ \psi_\alpha, \chi_\alpha \}$, a $SU(2)_N$ triplet 
of Grassmann--even ghost--for--ghost scalar fields 
$\phi^{ab} = \{ \phi, \tau, \bar{\phi} \}$, where $\tau$ plays the role 
of a Higgs field, and a $SU(2)_N$ doublet of Grassmann--odd scalar fields 
$\eta^a = \{ \lambda, \eta \}$, respectively.\break (Let us recall, that 
$\phi^{ab}$ is symmetric, $\phi^{ab} = \phi^{ba}$.) In order to close the 
topological superalgebra (see Eq. (2.4) below) it is necessary to introduce 
the bosonic auxiliary vector field $B_\alpha$. 
All the fields are in the adjoint representation and take their values in 
the Lie algebra $Lie(G)$ of some compact gauge group $G$.

The twisted action of this $D = 3$ super--BF (Casson or Euler character) model
\cite{24} with a $N_T = 2$ off--shell equivariantly nilpotent topological  
supersymmetry $Q^a$ is given by 
\begin{align}
\label{2.1}
S_{\rm BF} = \int d^3x\, {\rm tr} \Bigr\{&
\epsilon^{\alpha\beta\gamma} B_\gamma F_{\alpha\beta} +
\epsilon^{\alpha\beta\gamma} \epsilon_{ab} 
\psi_\gamma^a D_\alpha \psi_\beta^b - 
2 \epsilon_{ab} \eta^a D^\alpha \psi_\alpha^b -
2 \eta^a [ \phi_{ab}, \eta^b ]
\nonumber
\\
& - 2 \psi^{\alpha a} [ \phi_{ab}, \psi_\alpha^b ] + 
\phi_{ab} D^2 \phi^{ab} - [ \phi_{ab}, \phi_{cd} ] [ \phi^{ab}, \phi^{cd} ] - 
2 B^\alpha B_\alpha \Bigr\},
\end{align}
where 
$F_{\alpha\beta} = \partial_\alpha A_\beta - \partial_\beta A_\alpha + 
[ A_\alpha, A_\beta ]$ and $D_\alpha = \partial_\alpha + [ A_\alpha,~\cdot~ ]$
is the YM field strenght and the covariant derivative 
in the adjoint representation, respectively; 
$\epsilon_{\alpha\beta\gamma}$ is the totally antisymmetric Levi--Civita 
tensor in $D = 3$, $\epsilon_{123} = 1$, and $\epsilon_{ab}$ is the invariant 
tensor of the group $SU(2)_N$, $\epsilon_{12} = 1$. The internal index $a$, 
which labels the different $N_T = 2$ charges, is raised and lowered as 
follows: $\varphi_a = \varphi^b \epsilon_{ba}$ and 
$\varphi^a = \epsilon^{ab} \varphi_b$ with 
$\epsilon^{ac} \epsilon_{cb} = - \delta^a_b$. 

The action (\ref{2.1}) can be cast into the $Q^a$--exact form
\begin{align*}
S_{\rm BF} &= \hbox{\large$\frac{1}{2}$} \epsilon_{ab} Q^a Q^b X_{\rm BF},
\\
\intertext{where}
X_{\rm BF} &= S_{\rm CS} + \int d^3x\, {\rm tr} \Bigr\{
\epsilon_{ab} \psi^{\alpha a} \psi_\alpha^b + 
\epsilon_{ab} \eta^a \eta^b \Bigr\},
\\
\hbox{with}
\qquad
S_{\rm CS} &= \int d^3x\, {\rm tr} \Bigr\{
\epsilon^{\alpha\beta\gamma} ( A_\alpha \partial_\beta A_\gamma +
\hbox{\large$\frac{2}{3}$} A_\alpha A_\beta A_\gamma ) \Bigr\}
\end{align*}
being the Chern--Simons action. Let us notice, that the gauge boson 
$X_{\rm BF}$ is not uniquely specified, namely, substituting for
$\epsilon_{ab} \eta^a \eta^b$ the expression 
$\hbox{\large$\frac{2}{3}$} \epsilon_{ab} \phi_{cd} [ \phi^{ac}, \phi^{bd} ]$
gives the same action. Since the BF--term   
$\epsilon^{\alpha\beta\gamma} B_\gamma F_{\alpha\beta}$ in (\ref{2.1}) 
has an on--shell first--stage reducible gauge symmetry,
$\delta_G(\omega) B_\gamma = - D_\gamma \omega$, the gauge--fixing terms
can be derived also by the Batalin--Vilkovisky procedure \cite{13,28}. 
Let us notice, that the gauge--fixed action 
of the $D = 3$ super--BF theory can also be obtained by a dimensional 
reduction \cite{29} of the $N_T = 1$, $D = 4$ Donaldson--Witten theory 
\cite{3} which results by twisting the $N = 2$, $D = 4$ SYM (see Diagram). 

The off--shell equivariantly nilpotent topological shift symmetry $Q^a$ takes 
the form 
\begin{alignat}{2}
\label{2.2}
Q^a A_\alpha &= \psi_\alpha^a,
&\qquad&
\nonumber
\\
Q^a \phi^{bc} &= \hbox{\large$\frac{1}{2}$} \epsilon^{ab} \eta^{c} +
\hbox{\large$\frac{1}{2}$} \epsilon^{ac} \eta^{b},
&\qquad
Q^a \eta^b &= - \epsilon_{cd} [ \phi^{ac}, \phi^{bd} ],
\nonumber
\\
Q^a \psi_\alpha^b &= D_\alpha \phi^{ab} + \epsilon^{ab} B_\alpha,
&\qquad
Q^a B_\alpha &= - \hbox{\large$\frac{1}{2}$} D_\alpha \eta^a -
\epsilon_{cd} [ \phi^{ac}, \psi_\alpha^d ];
\end{alignat}
it agrees with that of Ref.~\cite{18,30} (up to the ubiquitous factor of 
$\hbox{\large$\frac{1}{2}$}$ in front of $\eta^a$) and \cite{16}. In addition,
by restricting to flat Euclidean space--time the action (\ref{2.1}) is 
invariant also under the following vector supersymmetry $\bar{Q}_\alpha^a$,
\begin{align}
\label{2.3}
&\bar{Q}_\alpha^a A_\beta = \delta_{\alpha\beta} \eta^a + 
\epsilon_{\alpha\beta\gamma} \psi^{\gamma a},
\nonumber
\\
&\bar{Q}_\alpha^a \phi^{bc} = - \hbox{\large$\frac{1}{2}$} 
\epsilon^{ab} \psi_\alpha^c - \hbox{\large$\frac{1}{2}$} 
\epsilon^{ac} \psi_\alpha^b,
\nonumber
\\
&\bar{Q}_\alpha^a \eta^b = D_\alpha \phi^{ab} + \epsilon^{ab} B_\alpha,
\nonumber
\\ 
&\bar{Q}_\alpha^a \psi_\beta^b = - \epsilon^{ab} F_{\alpha\beta} +
\epsilon^{ab} \epsilon_{\alpha\beta\gamma} B^\gamma - 
\epsilon_{\alpha\beta\gamma} D^\gamma \phi^{ab} +
\delta_{\alpha\beta} \epsilon_{cd} [ \phi^{ac}, \phi^{bd} ],
\nonumber
\\
&\bar{Q}_\alpha^a B_\beta = D_\alpha \psi_\beta^a - 
\hbox{\large$\frac{1}{2}$} D_\beta \psi_\alpha^a +
\epsilon_{\alpha\beta\gamma} D^\gamma \eta^a +
\epsilon_{cd} [ \phi^{ac}, \epsilon_{\alpha\beta\gamma} \psi^{\gamma d} - 
\delta_{\alpha\beta} \eta^d ]. 
\end{align} 
By a straightforward calculation it can be verified that the four 
supercharges $Q^a$ and $\bar{Q}_\alpha^a$, together with the generator 
$P_\alpha = i \partial_\alpha$ of space--time translations,
obey the following topological superalgebra:
\begin{align}
\label{2.4}
\{ Q^a, Q^b \} &= - 2 \delta_G(\phi^{ab}),
\nonumber
\\
\{ Q^a, \bar{Q}_\alpha^b \} &= \epsilon^{ab} ( 
- i P_\alpha + \delta_G(A_\alpha) ),
\nonumber
\\
\{ \bar{Q}_\alpha^a, \bar{Q}_\beta^b \} &\doteq - 2 \delta_{\alpha\beta} 
\delta_G(\phi^{ab}),
\end{align}
where the symbol $\doteq$ means that the corresponding relation is satisfied 
only on--shell, i.e., by taking into account the equation of motions. Since 
both the supersymmetries $Q^a$ and $\bar{Q}_\mu^a$ are realized nonlinearly, 
the superalgebra (\ref{2.4}) closes only modulo the field--dependent gauge 
transformations $\delta_G(\omega)$, $\omega = \{ A_\alpha, \phi^{ab} \}$,
which are defined by $\delta_G(\omega) A_\alpha = - D_\alpha \omega$ and
$\delta_G(\omega) \varphi = [ \omega, \varphi ]$, $\varphi = 
\{ \phi^{ab}, \eta^a, \psi_\alpha^a, B_\alpha \}$.
Let us emphasize that the form of the action (\ref{2.1}) is not completely
specified by the topological supersymmetry $Q^a$, i.e., it is not the most 
general action compatible with the gauge and the $Q^a$--invariance. 
Nevertheless, it turns out to be uniquely characterized by imposing the 
vector supersymmetry $\bar{Q}_\alpha^a$. The conditions
$Q^a S_{\rm BF} = \bar{Q}_\alpha^a S_{\rm BF} = 0$ fix all the relative
numerical coefficients of the action (\ref{2.1}), allowing, in particular,
for a single coupling constant.
\medskip

\noindent
{\it (B) The Blau--Thompson model}
\\
The second twist of $N = 4$ SYM in $D = 3$ consists in 
replacing $SU(2)_E \otimes SU(2)_N$ through its diagonal subgroup. 
This yields the novel
$N_T = 2$ topological twist introduced by Blau and Thompson \cite{16}. The
gauge multiplet $\{ A_\alpha, V_\alpha, \psi_\alpha^a, \bar{\eta}^a \}$ of  
this topological model is built up from the gauge field $A_\alpha$,
a bosonic vector field $V_\alpha$, a $SU(2)_R$ doublet of Grassmann--odd
topological ghost--antighost vector fields $\psi_\alpha^a = \{
\psi_\alpha, \chi_\alpha \}$ and a $SU(2)_R$ doublet of Grassmann--odd
scalar fields $\bar{\eta}^a = \{ \bar{\lambda}, \bar{\eta} \}$. In order to 
close the topological superalgebra (see Eq.~(\ref{2.8}) below) it is necessary
to introduce a further set of bosonic auxiliary fields, namely
two vector fields $B_\alpha$, $\bar{B}_\alpha$ and a scalar field $Y$, 
respectively.

The twisted action of this topological model is given by \cite{16}
\begin{align}
\label{2.5}
S_{\rm BT} = \hbox{\large$\frac{1}{2}$} \int d^3x\, {\rm tr} \Bigr\{&
- i \epsilon^{\alpha\beta\gamma} B_\gamma F_{\alpha\beta}(A + i V) - 
i \epsilon^{\alpha\beta\gamma} \epsilon_{ab} 
\psi_\gamma^a D_\alpha(A + i V) \psi_\beta^b - 
4 B^\alpha \bar{B}_\alpha  - 4 Y^2 
\nonumber
\\
& + i \epsilon^{\alpha\beta\gamma} \bar{B}_\gamma F_{\alpha\beta}(A - i V) - 
2 \epsilon_{ab} \bar{\eta}^a D^\alpha(A - i V) \psi_\alpha^b - 
4 Y D^\alpha(A) V_\alpha \Bigr\},
\end{align}
and can be rewritten as  sum of a BF--like topological term and a
$Q^a$--exact term,
\begin{equation*}
S_{\rm BT} = \hbox{\large$\frac{1}{2}$} \int d^3x\, {\rm tr} \Bigr\{
i \epsilon^{\alpha\beta\gamma} \bar{B}_\gamma F_{\alpha\beta}(A - i V) 
\Bigr\} + \hbox{\large$\frac{1}{2}$} \epsilon_{ab} Q^a Q^b X_{\rm BT},
\end{equation*}
with the gauge boson
\begin{equation*}
X_{\rm BT} = - \hbox{\large$\frac{1}{4}$} i S_{\rm CS}(A + i V) - 
\int d^3x\, {\rm tr} \Bigr\{ i \bar{B}^\alpha V_\alpha + 
\hbox{\large$\frac{1}{2}$} \epsilon_{ab} \bar{\eta}^a \bar{\eta}^b \Bigr\}.
\end{equation*}
Here, the Chern--Simons action $S_{\rm CS}(A + i V)$ is formed by the 
complexified gauge field $A_\alpha + i V_\alpha$. A striking, but somewhat 
unusual feature of this model is that there are no bosonic scalar fields and 
hence no underlying equivariant $Q^a$--cohomology (after dimensional 
reduction the three scalar fields are combined to form the vector field 
$V_\alpha$). Another special feature is that $A_\alpha - i V_\alpha$ is 
$Q^a$--invariant. Thus, as pointed out in \cite{16}, any gauge invariant 
functional of $A_\alpha - i V_\alpha$, constrained by 
$F_{\alpha\beta}(A - i V) = 0$, is a good observable 
(e.g., bosonic Wilson loops). Moreover, since this twisted model differs 
from the $D = 3$ super--BF model by an exchange of $SU(2)_R$ and $SU(2)_N$, 
in Ref. \cite{16} it has been speculated, that it can be regarded as 
providing a mirror description of the Casson model.  

Let us now give the transformation laws which leave the action (\ref{2.5})
invariant. The off--shell nilpotent topological supersymmetry 
$Q^a$ takes the form \cite{16}
\begin{alignat}{2}
\label{2.6}
&Q^a A_\alpha = \psi_\alpha^a,
&\qquad
&Q^a V_\alpha = - i \psi_\alpha^a, 
\nonumber
\\
&Q^a \psi_\alpha^b = 2 \epsilon^{ab} B_\alpha,
&\qquad
&Q^a \bar{\eta}^b = - 2 i \epsilon^{ab} Y,
\nonumber
\\
&Q^a B_\alpha = 0,
&\qquad
&Q^a \bar{B}_\alpha = - D_\alpha(A - i V) \bar{\eta}^a,
\nonumber
\\
&Q^a Y = 0,
&\qquad&
\end{alignat}
i.e., prior to the introduction of gauge ghosts, the $N_T = 2$ topological 
supersymmetry $Q^a$ is not equivariant, but rather strictly nilpotent. In 
addition, by restricting to flat Euclidean space--time, the action 
(\ref{2.5}) is left invariant under the following vector supersymmetry 
$Q_\alpha^a$, 
\begin{align}
\label{2.7}
&Q_\alpha^a A_\beta = \delta_{\alpha\beta} \bar{\eta}^a -
i \epsilon_{\alpha\beta\gamma} \psi^{\gamma a},
\nonumber
\\
&Q_\alpha^a \bar{\eta}^b = 2 \epsilon^{ab} \bar{B}_\alpha,
\nonumber
\\
&Q_\alpha^a \bar{B}_\beta = - i \epsilon_{\alpha\beta\gamma} 
D^\gamma(A + i V) \bar{\eta}^a,
\nonumber
\\
&Q_\alpha^a V_\beta = - i \delta_{\alpha\beta} \bar{\eta}^a +
\epsilon_{\alpha\beta\gamma} \psi^{\gamma a},
\nonumber
\\
&Q_\alpha^a \psi_\beta^b = - 2 \epsilon^{ab} F_{\alpha\beta}(A) - 
2 i \epsilon^{ab} D_\alpha(A) V_\beta +
2 i \epsilon^{ab} \epsilon_{\alpha\beta\gamma} \bar{B}^\gamma -
2 i \delta_{\alpha\beta} \epsilon^{ab} Y,
\nonumber
\\
&Q_\alpha^a B_\beta = 2 D_\alpha(A) \psi_\beta^a - 
D_\beta(A + i V) \psi_\alpha^a +
i \epsilon_{\alpha\beta\gamma} D^\gamma(A - i V) \bar{\eta}^a,
\nonumber
\\
&Q_\alpha^a Y = i D_\alpha(A - i V) \bar{\eta}^a.
\end{align} 
The scalar and the vector supercharges, $Q^a$ and $Q_\alpha^a$, together 
with the generator $P_\alpha$ of space--time translations, satisfy the 
following topological superalgebra: 
\begin{align}
\label{2.8}
\{ Q^a, Q^b \} &= 0,
\nonumber
\\
\{ Q^a, Q_\alpha^b \} &= \epsilon^{ab} ( 
- i P_\alpha + \delta_G(A_\alpha - i V_\alpha) ),
\nonumber
\\
\{ Q_\alpha^a, Q_\beta^b \} &\doteq 2 i \epsilon^{ab} 
\epsilon_{\alpha\beta\gamma} ( 
- i P^\gamma + \delta_G(A^\gamma - i V^\gamma) ).
\end{align}
As before, all relative numerical factors of the action (\ref{2.5}), except
for an overall unique coupling constant, are fixed by imposing the
requirements $Q^a S_{\rm BT} = Q_\alpha^a S_{\rm BT} = 0$.
\bigskip
\begin{flushleft}
{\large{\bf 3. $N_T = 4$ equivariant extension of Blau--Thompson model}}
\end{flushleft}
\bigskip
After having characterized the two possible topological twists of $N = 4$, 
$D = 3$ SYM let us turn to the question whether the topological model 
constructed by Blau and Thompson can be regarded as deformation of another 
one with underlying $N_T \geq 2$ equivariant cohomology. 
Under the requirement of preserving the number of topological 
supercharges, $N_T = 2$, we have not been able to find an appropiate 
equivariant extension of this model. However, if the condition $N_T = 2$ is 
relaxed, it is not difficult to show that this model, formally, can be 
recovered by deforming a cohomological theory with an extended $N_T = 4$ 
topological supersymmetry. Such a theory has an additional global symmetry 
group, which will be denoted by $\overline{SU(2)}_R$. In Section 4 
it will be shown that this cohomological theory coincides with the $N_T = 4$ 
topological twist of $N = 8$ SYM in $D = 3$.

The construction of the $N_T = 4$ equivariant extension of the Blau--Thompson
topological model is governed by the following strategy:
\\
First, after eliminating, by the use of the equations of motion, in the action
(\ref{2.5}) the auxiliary fields $B_\alpha$, $\bar{B}_\alpha$ and $Y$ the 
bosonic part of the resulting action involves the complexified gauge fields
$A_\alpha \pm i V_\alpha$. This bears a strong resemblance to the so--called
topological B--twist of $N = 4$, $D = 4$ SYM studied by Markus \cite{23} 
(recalling that $A_\alpha$, $V_\alpha$ and $Y$ are anti--hermitean). 
Hence, $\bar{B}_\alpha$ should be regarded as the anti--hermitean conjugate 
of $B_\alpha$. 
\\
Second, we introduce a $\overline{SU(2)}_R$ doublet of Grassmann--odd 
topological ghost--antighost vector fields 
$\bar{\psi}_\alpha^a = \{ \bar{\psi}_\alpha, \bar{\chi}_\alpha \}$ and 
a $\overline{SU(2)}_R$ doublet of Grassmann--odd scalar fields 
$\eta^a = \{ \lambda, \eta \}$, which should be regarded as the 
hermitean conjugate of $\psi_\alpha^a$ and $\bar{\eta}^a$, respectively. 
Then, we construct an $SU(2)_R \otimes \overline{SU(2)}_R$ invariant action 
by adding to (\ref{2.5}) appropiate $\bar{\psi}_\alpha^a$-- and 
$\eta^a$--dependent terms. 
\\
Third, we introduce a $SU(2)_R \otimes \overline{SU(2)}_R$ quartet of 
Grassmann--even ghost--for--ghost scalar fields $\zeta^{ab} = 
\{ \phi, \tau + i \rho, \tau - i \rho, \bar{\phi} \}$ and
$\bar{\zeta}^{ab} \equiv \zeta^{ba} =
\{ \phi, \tau - i \rho, \tau + i \rho, \bar{\phi} \}$, where
$\tau \pm i \rho$ plays the role of a complexified Higgs field. 
\\
Finally, we complete the action by adding suitable $\zeta^{ab}$-- and 
$\bar{\zeta}^{ab}$--dependent terms analogous to the $\phi^{ab}$--dependent 
terms in the action (\ref{2.1}) of the super--BF model.

Proceeding in that way one gets the following $N_T = 4$ equivariant 
extension of the action (\ref{2.5}), 
\begin{align}
\label{3.10}
S^{(N_T = 4)} = \hbox{\large$\frac{1}{2}$} \int d^3x\, {\rm tr} \Bigr\{&
- i \epsilon^{\alpha\beta\gamma} B_\gamma F_{\alpha\beta}(A + i V) - 
i \epsilon^{\alpha\beta\gamma} \epsilon_{ab} 
\psi_\gamma^a D_\alpha(A + i V) \psi_\beta^b
\nonumber
\\
& - 2 \epsilon_{ab} \bar{\eta}^a D^\alpha(A - i V) \psi_\alpha^b + 
2 \zeta_{ab} \{ \bar{\eta}^a, \eta^b \} +
2 \zeta_{ab} \{ \bar{\psi}^{\alpha a}, \psi_\alpha^b \} 
\phantom{\frac{1}{2}}
\nonumber
\\
& + \zeta_{ab} D^2(A + i V) \bar{\zeta}^{ab} -
[ \zeta_{ab}, \zeta_{cd} ] [ \bar{\zeta}^{ab}, \bar{\zeta}^{cd} ] - 
4 B^\alpha \bar{B}_\alpha
\phantom{\frac{1}{2}}
\nonumber
\\
& + i \epsilon^{\alpha\beta\gamma} \bar{B}_\gamma F_{\alpha\beta}(A - i V) + 
i \epsilon^{\alpha\beta\gamma} \epsilon_{ab} 
\bar{\psi}_\gamma^a D_\alpha(A - i V) \bar{\psi}_\beta^b
\phantom{\frac{1}{2}}
\nonumber
\\
& - 2 \epsilon_{ab} \eta^a D^\alpha(A + i V) \bar{\psi}_\alpha^b +
2 \bar{\zeta}_{ab} \{ \eta^a, \bar{\eta}^b \} +
2 \bar{\zeta}_{ab} \{ \psi^{\alpha a}, \bar{\psi}_\alpha^b \} 
\phantom{\frac{1}{2}}
\nonumber
\\
& + \bar{\zeta}_{ab} D^2(A - i V) \zeta^{ab} -
[ \bar{\zeta}_{ab}, \bar{\zeta}_{cd} ] [ \zeta^{ab}, \zeta^{cd} ] - 
4 Y D^\alpha(A) V_\alpha - 4 Y^2 \Bigr\},
\end{align}
which, by conctruction, is manifestly invariant under hermitean conjugation 
or, equivalently, under the discrete symmetry 
\begin{align}
\label{3.11}
( A_\alpha, V_\alpha, B_\alpha, \bar{B}_\alpha, Y )
&\rightarrow 
( A_\alpha, - V_\alpha, - \bar{B}_\alpha, - B_\alpha, - Y ), 
\nonumber
\\
( \psi_\alpha^a, \bar{\psi}_\alpha^a, \eta^a, \bar{\eta}^a,
\zeta^{ab}, \bar{\zeta}^{ab} )
&\rightarrow 
( i \bar{\psi}_\alpha^a, - i \psi_\alpha^a, i \bar{\eta}^a, - i \eta^a,
\bar{\zeta}^{ab}, \zeta^{ab} ),
\end{align} 
exchanging the scalar and the vector supercharges with their conjugate ones. 
Thus, the underlying equivariant cohomology should be a 
$N_T = 4$ supersymmetry. This is indeed the case. By an explicit calculation 
one establishes that the action (\ref{3.10}) is invariant under the following 
off--shell equivariantly nilpotent topological supersymmetry $Q^a$,
\begin{alignat}{2}
\label{3.12}
&Q^a A_\alpha = \psi_\alpha^a,
&\qquad
&Q^a V_\alpha = - i \psi_\alpha^a, 
\nonumber
\\
&Q^a \zeta^{bc} = \epsilon^{ac} \eta^b,
&\qquad
&Q^a \bar{\zeta}^{bc} = \epsilon^{ab} \eta^c,
\nonumber
\\
&Q^a \psi_\alpha^b = 2 \epsilon^{ab} B_\alpha,
&\qquad
&Q^a \bar{\psi}_\alpha^b = 2 D_\alpha(A - i V) \bar{\zeta}^{ab},
\nonumber
\\
&Q^a \eta^b = 0,
&\qquad
&Q^a \bar{\eta}^b = - 2 i \epsilon^{ab} Y - 
2 \epsilon_{cd} [ \bar{\zeta}^{ac}, \bar{\zeta}^{bd} ],
\nonumber
\\
&Q^a B_\alpha = 0,
&\qquad
&Q^a \bar{B}_\alpha = - D_\alpha(A - i V) \bar{\eta}^a - 
2 \epsilon_{cd} [ \bar{\zeta}^{ac}, \bar{\psi}_\alpha^d ],
\nonumber
\\
&Q^a Y = i \epsilon_{cd} [ \bar{\zeta}^{ac}, \eta^d ],
&\qquad&
\\
\intertext{which, formally, may be regarded as deformation of the topological 
supersymmetry displayed in (\ref{2.6}). This deformation is, of course,
topological non--trivial since some of the fields, namely $\eta^a$,
$\bar{\psi}_\alpha^a$, $\zeta^{ab}$ and $\bar{\zeta}^{ab}$, must be deformed 
equal to zero. In addition, applying the discrete symmetry (\ref{3.11}) on 
$Q^a$, which maps $Q^a$ to $i \bar{Q}_\alpha^a$, one gets a further one, 
namely the conjugate topological supersymmetry $\bar{Q}^a$,} 
\label{3.13}
&\bar{Q}^a A_\alpha = \bar{\psi}_\alpha^a,
&\qquad
&\bar{Q}^a V_\alpha = i \bar{\psi}_\alpha^a, 
\nonumber
\\
&\bar{Q}^a \bar{\zeta}^{bc} = \epsilon^{ac} \bar{\eta}^b,
&\qquad
&\bar{Q}^a \zeta^{bc} = \epsilon^{ab} \bar{\eta}^c,
\nonumber
\\
&\bar{Q}^a \bar{\psi}_\alpha^b = 2 \epsilon^{ab} \bar{B}_\alpha,
&\qquad
&\bar{Q}^a \psi_\alpha^b = 2 D_\alpha(A + i V) \zeta^{ab},
\nonumber
\\
&\bar{Q}^a \bar{\eta}^b = 0,
&\qquad
&\bar{Q}^a \eta^b = 2 i \epsilon^{ab} Y - 
2 \epsilon_{cd} [ \zeta^{ac}, \zeta^{bd} ],
\nonumber
\\
&\bar{Q}^a \bar{B}_\alpha = 0,
&\qquad
&\bar{Q}^a B_\alpha = - D_\alpha(A + i V) \eta^a - 
2 \epsilon_{cd} [ \zeta^{ac}, \psi_\alpha^d ],
\nonumber
\\
&\bar{Q}^a Y = - i \epsilon_{cd} [ \zeta^{ac}, \bar{\eta}^d ],
&\qquad&
\end{alignat}
proving, as promised, that the action (\ref{3.10}) actually possesses a 
$N_T = 4$ topological supersymmetry. 

By restricting to flat Euclidean space--time one can convince oneself
by a rather lengthy calculation that the action (\ref{3.10}) is also
invariant under the following vector supersymmetry $\bar{Q}_\alpha^a$:
\begin{align}
\label{3.14}
&\bar{Q}_\alpha^a A_\beta = \delta_{\alpha\beta} \eta^a +
i \epsilon_{\alpha\beta\gamma} \bar{\psi}^{\gamma a},
\nonumber
\\
&\bar{Q}_\alpha^a \zeta^{bc} = - \epsilon^{ab} \psi_\alpha^c,
\nonumber
\\
&\bar{Q}_\alpha^a \psi_\beta^b = - 2 i \epsilon_{\alpha\beta\gamma}
D^\gamma(A - i V) \zeta^{ab},
\nonumber
\\
&\bar{Q}_\alpha^a \eta^b = 2 \epsilon^{ab} B_\alpha,
\nonumber
\\
&\bar{Q}_\alpha^a B_\beta = i \epsilon_{\alpha\beta\gamma} 
D^\gamma(A - i V) \eta^a,
\nonumber
\\
&\bar{Q}_\alpha^a V_\beta = i \delta_{\alpha\beta} \eta^a +
\epsilon_{\alpha\beta\gamma} \bar{\psi}^{\gamma a},
\nonumber
\\
&\bar{Q}_\alpha^a \bar{\zeta}^{bc} = - \epsilon^{ac} \psi_\alpha^b,
\nonumber
\\
&\bar{Q}_\alpha^a \bar{\psi}_\beta^b = - 2 \epsilon^{ab} F_{\alpha\beta}(A) + 
2 i \epsilon^{ab} D_\alpha(A) V_\beta - 
2 i \epsilon^{ab} \epsilon_{\alpha\beta\gamma} B^\gamma +
2 \delta_{\alpha\beta} \epsilon_{cd} [ \zeta^{ac}, \zeta^{bd} ] +
2 i \delta_{\alpha\beta} \epsilon^{ab} Y,
\nonumber
\\
&\bar{Q}_\alpha^a \bar{\eta}^b = 2 D_\alpha(A + i V) \zeta^{ab},
\nonumber
\\
&\bar{Q}_\alpha^a \bar{B}_\beta = 2 D_\alpha(A) \bar{\psi}_\beta^a - 
D_\beta(A - i V) \bar{\psi}_\alpha^a -
i \epsilon_{\alpha\beta\gamma} D^\gamma(A + i V) \eta^a -
2 \epsilon_{cd} [ \zeta^{ac}, \delta_{\alpha\beta} \bar{\eta}^d +
i \epsilon_{\alpha\beta\gamma} \psi^{\gamma d} ].
\nonumber
\\
&\bar{Q}_\alpha^a Y = - i D_\alpha(A + i V) \eta^a -
i \epsilon_{cd} [ \zeta^{ac}, \psi_\alpha^d ],
\\
\intertext{Combining $\bar{Q}_\alpha^a$ with the discrete symmetry 
(\ref{3.11}), which maps $\bar{Q}_\alpha^a$ into $i Q_\alpha^a$, gives rise 
to the conjugate vector supersymmetry $Q_\alpha^a$,}
\label{3.15}
&Q_\alpha^a A_\beta = \delta_{\alpha\beta} \bar{\eta}^a -
i \epsilon_{\alpha\beta\gamma} \psi^{\gamma a},
\nonumber
\\
&Q_\alpha^a \bar{\zeta}^{bc} = - \epsilon^{ab} \bar{\psi}_\alpha^c,
\nonumber
\\
&Q_\alpha^a \bar{\psi}_\beta^b = 2 i \epsilon_{\alpha\beta\gamma}
D^\gamma(A + i V) \bar{\zeta}^{ab},
\nonumber
\\
&Q_\alpha^a \bar{\eta}^b = 2 \epsilon^{ab} \bar{B}_\alpha,
\nonumber
\\
&Q_\alpha^a \bar{B}_\beta = - i \epsilon_{\alpha\beta\gamma} 
D^\gamma(A + i V) \bar{\eta}^a,
\nonumber
\\
&Q_\alpha^a V_\beta = - i \delta_{\alpha\beta} \bar{\eta}^a +
\epsilon_{\alpha\beta\gamma} \psi^{\gamma a},
\nonumber
\\
&Q_\alpha^a \zeta^{bc} = - \epsilon^{ac} \bar{\psi}_\alpha^b,
\nonumber
\\
&Q_\alpha^a \psi_\beta^b = - 2 \epsilon^{ab} F_{\alpha\beta}(A) - 
2 i \epsilon^{ab} D_\alpha(A) V_\beta + 
2 i \epsilon^{ab} \epsilon_{\alpha\beta\gamma} \bar{B}^\gamma +
2 \delta_{\alpha\beta} \epsilon_{cd} [ \bar{\zeta}^{ac}, \bar{\zeta}^{bd} ] -
2 i \delta_{\alpha\beta} \epsilon^{ab} Y,
\nonumber
\\
&Q_\alpha^a \eta^b = 2 D_\alpha(A - i V) \bar{\zeta}^{ab},
\nonumber
\\
&Q_\alpha^a B_\beta = 2 D_\alpha(A) \psi_\beta^a - 
D_\beta(A + i V) \psi_\alpha^a +
i \epsilon_{\alpha\beta\gamma} D^\gamma(A - i V) \bar{\eta}^a -
2 \epsilon_{cd} [ \bar{\zeta}^{ac}, \delta_{\alpha\beta} \eta^d -
i \epsilon_{\alpha\beta\gamma} \bar{\psi}^{\gamma d} ],
\nonumber
\\
&Q_\alpha^a Y = i D_\alpha(A - i V) \bar{\eta}^a +
i \epsilon_{cd} [ \bar{\zeta}^{ac}, \bar{\psi}_\alpha^d ],
\end{align} 
which, formally, may be regarded as deformation of the vector 
supersymmetry given in (\ref{2.7}). The eight supercharges $Q^a$, $\bar{Q}^a$, 
$\bar{Q}_\alpha^a$ and $Q_\alpha^a$, together with the generator $P_\alpha$ 
of space--time translations, obey the following topological 
superalgebra,
\begin{alignat}{2}
\label{3.16}
\{ Q^a, Q^b \} &= 0,
&\qquad
\{ Q^a, \bar{Q}^b \} &= - 4 \delta_G(\bar{\zeta}^{ab}),
\nonumber
\\
\{ \bar{Q}^a, \bar{Q}^b \} &= 0,
&\qquad
\{ \bar{Q}^a, Q^b \} &= - 4 \delta_G(\zeta^{ab}),
\nonumber
\\
\{ Q^a, \bar{Q}_\alpha^b \} &= 0,
&\qquad
\{ Q^a, Q_\alpha^b \} &= 2 \epsilon^{ab} ( 
- i P_\alpha + \delta_G(A_\alpha - i V_\alpha) ),
\nonumber
\\
\{ \bar{Q}^a, Q_\alpha^b \} &= 0,
&\qquad
\{ \bar{Q}^a, \bar{Q}_\alpha^b \} &= 2 \epsilon^{ab} (
- i P_\alpha + \delta_G(A_\alpha + i V_\alpha) ),
\nonumber
\\
\{ Q_\alpha^a, \bar{Q}_\beta^b \} &\doteq - 4 \epsilon^{ab} 
\delta_{\alpha\beta} \delta_G(\bar{\zeta}^{ab}),
&\qquad
\{ Q_\alpha^a, Q_\beta^b \} &\doteq 2 i \epsilon^{ab}
\epsilon_{\alpha\beta\gamma} ( 
- i P^\gamma + \delta_G(A^\gamma - i V^\gamma) ),
\nonumber
\\
\{ \bar{Q}_\alpha^a, Q_\beta^b \} &\doteq - 4 \epsilon^{ab} 
\delta_{\alpha\beta} \delta_G(\zeta^{ab}),
&\qquad
\{ \bar{Q}_\alpha^a, \bar{Q}_\beta^b \} &\doteq - 2 i \epsilon^{ab}
\epsilon_{\alpha\beta\gamma} ( 
- i P^\gamma + \delta_G(A^\gamma + i V^\gamma) ),
\end{alignat}
which is just the $N_T = 4$ equivariant extension of the topological
superalgebra (\ref{2.8}). 

In summary, so far we have seen that the novel $N_T = 2$ topological twist 
of $N = 4$, $D = 3$ SYM can be regarded, formally, as restriction of a 
$N_T = 4$ topological gauge theory.
\bigskip
\begin{flushleft}
{\large{\bf 4. Dimensional reduction of $N_T = 2$, $D = 4$ topological 
YM theory to $D = 3$}}
\end{flushleft}
\bigskip
Now, our aim is to show that the $N_T = 4$ equivariant extension of the  
Blau--Thompson model introduced in Section 3 is precisely one of the two
essentially different topological twists of $N = 8$, $D = 4$ SYM which, on the
other hand, is just the dimensional reduction of either of the two $N_T = 2$,
$D = 4$ models to $D = 3$, namely, either the $A$--model \cite{18,22} or the 
$B$--model \cite{23}. However, before proceeding, let us briefly recall the 
two possible topological twists of $N = 8$, $D = 3$ SYM and its relation to 
the three essentially different topological twists of the 
$N = 4$, $D = 4$ SYM. 

The $N = 8$, $D = 3$ SYM obtains by dimensional reduction of 
$N = 1$, $D = 10$ SYM, either directly or via $N = 4$, $D = 4$ SYM, to 
$D = 3$ (cf.,~lower half of the Diagram). The global 
symmetry group of $N = 8$, $D = 3$ SYM is $SU(2)_E \otimes Spin(7)$.  
In decomposing $Spin(7)$, with respect to the twist procedure, some 
restrictions have to be required \cite{16}:
\\
First, the twisted theory should contain at least one scalar topological 
supercharge.
\\
Second, among the spinor representations of $Spin(7)$ no ones with spin 
$\geq 2$ should appear.
\\
Third, for a full topological twist only half--integral spins should appear 
among the spinor representations. 
\\
Under these restrictions $Spin(7)$ decomposes as
$Spin(7) \rightarrow SU(2)_R \otimes \overline{SU(2)}_R \otimes SU(2)_N$,
so that the maximal residual global symmetry group is
$SU(2)_R \otimes \overline{SU(2)}_R$. 

Now, choosing the diagonal subgroup of $SU(2)_E \otimes SU(2)_N$ one gets 
a twisted theory with an underlying $N_T = 4$ equivariant cohomology. Below, 
it will be shown that the action of this model is precisely the one given 
in (\ref{3.10}). 

The other topological twist is obtained by taking the diagonal subgroup 
of $SU(2)_E \otimes SU(2)_R$ and gives rise to a $N_T = 2$ theory with 
global symmetry group $SU(2)_N \otimes \overline{SU(2)}_R$. The action of 
that theory will be explicitly constructed in Section 5. 

Furthermore, there are three topological twists of $N = 4$, $D = 4$ SYM, 
namely the A--model, which is the $N_T = 2$ equivariant extension of the 
$N_T = 1$, $D = 4$ Donaldson--Witten model \cite{3}, the B--model, 
which formally can be regarded as a deformation \cite{16} 
of the $N_T = 2$, $D = 4$ super--BF model \cite{28}, and the half--twisted
$N_T = 1$, $D = 4$ model \cite{22}, the Donaldson--Witten theory coupled to
a spinoral hypermultiplet. Therefore, one might have expected that there are
at least three topological twists of $N = 8$, $D = 3$ SYM. But, as pointed out 
in \cite{16}, the dimensional reduction of either of the two 
$N_T = 2$, $D = 4$ theories, i.e., the A-- and the B--model, give rise 
to equivalent $D = 3$ topological gauge theories, so that, under the 
above--mentioned restrictions, there are only two different twists.

Now we want to show that the dimensional reduction of the $N_T = 2$, $D = 4$ 
Yamron--Vafa--Witten theory with global symmetry group $SU(2)_R$, 
i.e.,~the A--model, leads precisely to the action given in (\ref{3.10}). 
The gauge multiplet of this theory consists of the gauge field $A_\mu$, a 
Grassmann--even self--dual tensor field $M_{\mu\nu}$, a $SU(2)_R$ doublet
of Grassmann--odd self--dual ghost--antighost tensor fields 
$\chi_{\mu\nu}^a = \{ \psi_{\mu\nu}, \chi_{\mu\nu} \}$, a 
$SU(2)_R$ doublet of Grassmann--odd ghost--antighost vector fields
$\psi_\mu^a = \{ \psi_\mu, \chi_\mu \}$, a $SU(2)_R$ doublet 
of Grassmann--odd scalar fields $\eta^a = \{ \lambda, \eta \}$, and a
$SU(2)_R$ triplet of Grassmann--even ghost--for--ghost complex scalars 
$\phi^{ab} = \{ \phi, \tau, \bar{\phi} \}$. For the closure of the 
topological superalgebra (see Eq. (\ref{4.20}) below) it is necessary
to introduce a set of bosonic auxiliary fields, namely the self--dual tensor 
field $G_{\mu\nu}$ and the vector field $H_\mu$. All the fields are in the 
adjoint representation and take their values in the Lie algebra $Lie(G)$ of 
some compact gauge group $G$.

The action of the Yamron--Vafa--Witten model, with an $N_T = 2$ off--shell 
equivariantly nilpotent topological shift symmetry $Q^a$, is given by 
\cite{18,22} (see also \cite{31})
\begin{align}
\label{4.17}
S_{\rm YVW} = \int d^4x\, {\rm tr} \Bigr\{&
G^{\mu\nu} F_{\mu\nu} - \hbox{\large$\frac{1}{4}$} 
G^{\mu\nu} [ M_\mu^{~\lambda}, M_{\nu\lambda} ] + 
2 \epsilon_{ab} \chi^{\mu\nu a} D_\mu \psi_\nu^b - 
\hbox{\large$\frac{1}{4}$} \epsilon_{ab} 
\chi^{\mu\nu a} [ M_\mu^{~\lambda}, \chi_{\nu\lambda}^b ]
\nonumber
\\
& - \epsilon_{ab} \psi^{\mu a} [ M_{\mu\nu}, \psi^{\nu b} ] - 
2 \epsilon_{ab} \eta^a D^\mu \psi_\mu^b - 
\hbox{\large$\frac{1}{2}$} \epsilon_{ab} 
\eta^a [ M^{\mu\nu}, \chi_{\mu\nu}^b ] -
\hbox{\large$\frac{1}{2}$} G^{\mu\nu} G_{\mu\nu}
\phantom{\frac{1}{2}}
\nonumber
\\
& + 2 \phi_{ab} \{ \eta^a, \eta^b \} + 
2 \phi_{ab} \{ \psi^{\mu a}, \psi_\mu^b \} + 
\hbox{\large$\frac{1}{2}$} \phi_{ab} \{ \chi^{\mu\nu a}, \chi_{\mu\nu}^b \} +
\phi_{ab} D^2 \phi^{ab} 
\phantom{\frac{1}{2}}
\nonumber
\\
& - \hbox{\large$\frac{1}{4}$} 
[ \phi_{ab}, M^{\mu\nu} ] [ \phi^{ab}, M_{\mu\nu} ] -
[ \phi_{ab}, \phi_{cd} ] [ \phi^{ab}, \phi^{cd} ] + 
2 H^\mu D^\nu M_{\mu\nu} - 2 H^\mu H_\mu \Bigr\},
\end{align}
and can be cast into the $Q^a$--exact form
\begin{equation*}
S_{\rm YVW} = \hbox{\large$\frac{1}{2}$} \epsilon_{ab} Q^a Q^b X_{\rm YVW},
\end{equation*}
with the gauge boson
\begin{equation*}
X_{\rm YVW} = \int d^4x\, {\rm tr} \Bigr\{
M^{\mu\nu} F_{\mu\nu} -
\hbox{\large$\frac{1}{12}$} M^{\mu\nu} [ M_\mu^{~\lambda}, M_{\nu\lambda} ] - 
\hbox{\large$\frac{1}{2}$} G^{\mu\nu} M_{\mu\nu} + 
\epsilon_{ab} \psi^{\mu a} \psi_\mu^b + \epsilon_{ab} \eta^a \eta^b \Bigr\}.
\end{equation*}
The complete set of symmetry transformations which fix all the relative 
numerical factors in (\ref{4.17}), except for an overall coupling 
constant, is given by the topological supersymmetry $Q^a$,
\begin{alignat}{2}
\label{4.18}
&Q^a A_\mu = \psi_\mu^a,
&\qquad
&Q^a M_{\mu\nu} = \chi_{\mu\nu}^a,
\nonumber
\\
&Q^a \phi^{bc} = \hbox{\large$\frac{1}{2}$} \epsilon^{ab} \eta^c +
\hbox{\large$\frac{1}{2}$} \epsilon^{ac} \eta^b,
&\qquad
&Q^a \eta^b = - \epsilon_{cd} [ \phi^{ac}, \phi^{bd} ],
\nonumber
\\
&Q^a \psi_\mu^b = D_\mu \phi^{ab} + \epsilon^{ab} H_\mu,
&\qquad
&Q^a H_\mu = - \hbox{\large$\frac{1}{2}$} D_\mu \eta^a -
\epsilon_{cd} [ \phi^{ac}, \psi_\mu^d ],
\nonumber
\\
&Q^a \chi_{\mu\nu}^b = [ M_{\mu\nu}, \phi^{ab} ] + 
\epsilon^{ab} G_{\mu\nu},
&\qquad
&Q^a G_{\mu\nu} = - \hbox{\large$\frac{1}{2}$} [ M_{\mu\nu}, \eta^a ] -
\epsilon_{cd} [ \phi^{ac}, \chi_{\mu\nu}^d ],
\end{alignat}
and by the vector supersymmetry $\bar{Q}_\mu^a$,
\begin{align}
\label{4.19}
&\bar{Q}_\mu^a A_\nu = \delta_{\mu\nu} \eta^a + \chi_{\mu\nu}^a,
\nonumber
\\ 
&\bar{Q}_\mu^a M_{\rho\sigma} = - \delta_{\mu[\rho} \psi_{\sigma]}^a -
\epsilon_{\mu\nu\rho\sigma} \psi^{\nu a},
\nonumber
\\
&\bar{Q}_\mu^a \phi^{bc} = - \hbox{\large$\frac{1}{2}$} 
\epsilon^{ab} \psi_\mu^c - \hbox{\large$\frac{1}{2}$} 
\epsilon^{ac} \psi_\mu^b,  
\nonumber
\\
&\bar{Q}_\mu^a \eta^b = D_\mu \phi^{ab} + \epsilon^{ab} H_\mu,
\nonumber
\\
&\bar{Q}_\mu^a \psi_\nu^b = - \epsilon^{ab} F_{\mu\nu} +
\delta_{\mu\nu} \epsilon_{cd} [ \phi^{ac}, \phi^{bd} ] +
\epsilon^{ab} G_{\mu\nu} - [ M_{\mu\nu}, \phi^{ab} ],
\nonumber
\\
&\bar{Q}_\mu^a H_\nu = D_\mu \psi_\nu^a - 
\hbox{\large$\frac{1}{2}$} D_\nu \psi_\mu^a +
\epsilon_{cd} [ \phi^{ac}, \chi_{\mu\nu}^d - \delta_{\mu\nu} \eta^d ] +
[ M_{\mu\nu}, \eta^a ], 
\nonumber
\\
&\bar{Q}_\mu^a \chi_{\rho\sigma}^b = \delta_{\mu[\rho} D_{\sigma]} \phi^{ab} +
\epsilon_{\mu\nu\rho\sigma} D^\nu \phi^{ab} -
\epsilon^{ab} \delta_{\mu[\rho} H_{\sigma]} -
\epsilon^{ab} \epsilon_{\mu\nu\rho\sigma} H^\nu - 
\epsilon^{ab} D_\mu M_{\rho\sigma},
\nonumber
\\
&\bar{Q}_\mu^a G_{\rho\sigma} = D_\mu \chi_{\rho\sigma}^a -
\delta_{\mu[\rho} D_{\sigma]} \eta^a -
\epsilon_{\mu\nu\rho\sigma} D^\nu \eta^a -
\epsilon_{cd} [ \phi^{ac}, \delta_{\mu[\rho} \psi_{\sigma]}^d +
\epsilon_{\mu\nu\rho\sigma} \psi^{\nu d} ] +
\hbox{\large$\frac{1}{2}$} [ \psi_\mu^a, M_{\rho\sigma} ], 
\end{align} 
which, together with the space--time translations $P_\mu$, obey the 
topological superalgebra
\begin{align}
\label{4.20}
\{ Q^a, Q^b \} &= - 2 \delta_G(\phi^{ab}),
\nonumber
\\
\{ Q^a, \bar{Q}_\mu^b \} &= \epsilon^{ab} ( - i P_\mu + \delta_G(A_\mu) ),
\nonumber
\\
\{ \bar{Q}_\mu^a, \bar{Q}_\nu^b \} &\doteq - 2 \delta_{\mu\nu} 
\delta_G(\phi^{ab}) - \epsilon^{ab} \delta_G(M_{\mu\nu}).
\end{align}

Our next goal is to show how the action (\ref{3.10}) is obtained from 
(\ref{4.17}) by a dimensional reduction. To this end we perform a
$(3 + 1)$--decomposition of the action (\ref{4.17}), i.e., we split the
space--time coordinates into $x^\mu = \{ x^\alpha, x^4 \}$, $\alpha = 1,2,3$,
where $x^\alpha$ and $x^4$ denote the spatial and the temporal 
part, respectively. As a next step, we assume that no field depends on 
$x^4$, i.e., $\partial_4 = 0$, so that the integration over $x^4$ factors out 
and, therefore, can be ignored. Furthermore, we rename the temporal part 
of $A_\mu$, $\psi_\mu^a$ and $H_\mu$, according to
\begin{equation*}
A_4 = 2 \rho,
\qquad 
\psi_4^a = \bar{\eta}^a,
\qquad
H_4 = Y,
\end{equation*}
reserving the notation $A_\alpha$ and $\psi_\alpha^a$ for the spatial
part of $A_\mu$ and $\psi_\mu^a$, and identify $M_{\mu\nu}$, $\chi_{\mu\nu}$, 
$G_{\mu\nu}$ and the spatial part of $H_\mu$ with
\begin{gather*}
M_{\alpha\beta} = \epsilon_{\alpha\beta\gamma} V^\gamma,
\qquad
M_{\alpha 4} = V_\alpha,
\qquad
\chi_{\alpha\beta}^a = \epsilon_{\alpha\beta\gamma} \bar{\psi}^{\gamma a},
\qquad
\chi_{\alpha 4}^a = \bar{\psi}_\alpha^a,
\\
G_{\alpha\beta} = \epsilon_{\alpha\beta\gamma} \bar{B}^\gamma +
\epsilon_{\alpha\beta\gamma} D^\gamma \rho,
\qquad
G_{\alpha 4} = \bar{B}_\alpha + D_\alpha \rho,
\qquad
H_\alpha = B_\alpha - [ V_\alpha, \rho ],
\end{gather*}
respectively. Here, the $\rho$--dependent shifts in $G_{\alpha\beta}$,
$G_{\alpha 4}$ and $H_\alpha$ ensure that after carrying out the dimensional 
reduction the hermitean conjugate $\bar{Q}^a$ of the scalar supercharge 
$Q^a$ coincides with the temporal part $\bar{Q}_4^a$ of the vector 
supercharge $\bar{Q}_\mu^a$, i.e., $\bar{Q}^a = \bar{Q}_4^a$.
Then, after squeezing (\ref{4.17}) to $D = 3$, we arrive at the following 
reduced action
\begin{align}
\label{4.21}
S^{(N_T = 4)} = \int d^3x\, {\rm tr} \Bigr\{&
\epsilon^{\alpha\beta\gamma} \bar{B}_\gamma F_{\alpha\beta} -
\epsilon^{\alpha\beta\gamma} \bar{B}_\gamma [ V_\alpha, V_\beta ] - 
2 \bar{B}^\alpha \bar{B}_\alpha + 
2 \epsilon_{ab} \epsilon^{\alpha\beta\gamma} 
\bar{\psi}_\gamma^a D_\alpha \psi_\beta^b
\nonumber
\\
& + \epsilon_{ab} \epsilon^{\alpha\beta\gamma} V_\gamma 
( \{ \psi_\alpha^a, \psi_\beta^b \} - 
\{ \bar{\psi}_\alpha^a, \bar{\psi}_\beta^b \} ) + 
2 \epsilon^{\alpha\beta\gamma} B_\gamma D_\alpha V_\beta - 
2 B^\alpha B_\alpha
\phantom{\frac{1}{2}}
\nonumber
\\
& - 2 \epsilon_{ab} ( \eta^a D^\alpha \psi_\alpha^b +
\bar{\eta}^a D^\alpha \bar{\psi}_\alpha^b ) + 
2 \epsilon_{ab} V^\alpha ( \{ \eta^a, \bar{\psi}_\alpha^b \} -
\{ \bar{\eta}^a, \psi_\alpha^b \} )
\phantom{\frac{1}{2}}
\nonumber
\\
& + 2 \phi_{ab} ( \{ \eta^a, \eta^b \} +
\{ \psi^{\alpha a}, \psi_\alpha^b \} ) + 
4 \rho \epsilon_{ab} ( \{ \eta^a, \bar{\eta}^b \} - 
\{ \psi^{\alpha a}, \bar{\psi}_\alpha^b \} )
\phantom{\frac{1}{2}}
\nonumber
\\
& + 2 \phi_{ab} ( \{ \bar{\eta}^a, \bar{\eta}^b \} +
\{ \bar{\psi}^{\alpha a}, \bar{\psi}_\alpha^b \} ) + 
\phi_{ab} D^2 \phi^{ab} + 2 \rho D^2 \rho - 
[ V^\alpha, \phi_{ab} ] [ V_\alpha, \phi^{ab} ]
\phantom{\frac{1}{2}}
\nonumber
\\
& - 2 [ V^\alpha, \rho ] [ V_\alpha, \rho ] -
[ \phi_{ab}, \phi_{cd} ] [ \phi^{ab}, \phi^{cd} ] -
4 [ \rho, \phi_{ab} ] [ \rho, \phi^{ab} ] - 
2 Y D^\alpha V_\alpha - 2 Y^2 \Bigr\}
\end{align}
which is manifestly invariant under the discrete symmetry 
\begin{align}
\label{4.22}
( A_\alpha, V_\alpha, B_\alpha, \bar{B}_\alpha, Y )
&\rightarrow 
( A_\alpha, - V_\alpha, - B_\alpha, \bar{B}_\alpha, - Y ), 
\nonumber
\\
( \psi_\alpha^a, \bar{\psi}_\alpha^a, \eta^a, \bar{\eta}^a,
\phi^{ab}, \rho )
&\rightarrow 
( \bar{\psi}_\alpha^a, \psi_\alpha^a, \bar{\eta}^a, \eta^a,
\phi^{ab}, - \rho ),
\end{align} 
exhibiting that the global symmetry group is actually 
$SU(2)_R \otimes \overline{SU(2)}_R$.

Next, after squeezing (\ref{4.18}) and (\ref{4.19}) to $D = 3$, for the 
transformation laws generated by the scalar and vector supercharges, 
$Q^a$ and $\bar{Q}_\alpha^a$, we get
\begin{align}
\label{4.23}
&Q^a A_\alpha = \psi_\alpha^a,
\nonumber
\\
&Q^a \phi^{bc} = \hbox{\large$\frac{1}{2}$} \epsilon^{ab} \eta^c +
\hbox{\large$\frac{1}{2}$} \epsilon^{ac} \eta^b,
\nonumber
\\
&Q^a \eta^b = - \epsilon_{cd} [ \phi^{ac}, \phi^{bd} ],
\nonumber
\\
&Q^a \psi_\alpha^b = D_\alpha \phi^{ab} - \epsilon^{ab} [ V_\alpha, \rho ] + 
\epsilon^{ab} B_\alpha,
\nonumber
\\
&Q^a B_\alpha = - \hbox{\large$\frac{1}{2}$} D_\alpha \eta^a +
\hbox{\large$\frac{1}{2}$} [ V_\alpha, \bar{\eta}^a ] -
\epsilon_{cd} [ \phi^{ac}, \psi_\alpha^d ] -
[ \rho, \bar{\psi}_\alpha^a ],
\nonumber
\\
&Q^a V_\alpha = \bar{\psi}_\alpha^a,
\nonumber
\\
&Q^a \rho = \hbox{\large$\frac{1}{2}$} \bar{\eta}^a,
\nonumber
\\
&Q^a \bar{\eta}^b = 2 [ \rho, \phi^{ab} ] + \epsilon^{ab} Y,
\nonumber
\\
&Q^a \bar{\psi}_\alpha^b = [ V_\alpha, \phi^{ab} ] + 
\epsilon^{ab} D_\alpha \rho + \epsilon^{ab} \bar{B}_\alpha,
\nonumber
\\
&Q^a \bar{B}_\alpha = - \hbox{\large$\frac{1}{2}$} D_\alpha \bar{\eta}^a - 
\hbox{\large$\frac{1}{2}$} [ V_\alpha, \eta^a ] -
\epsilon_{cd} [ \phi^{ac}, \bar{\psi}_\alpha^d ] + [ \rho, \psi_\alpha^a ],
\nonumber
\\
&Q^a Y = - [ \rho, \eta^a ] - \epsilon_{cd} [ \phi^{ac}, \bar{\eta}^d ]
\\
\intertext{and}
\label{4.24}
&\bar{Q}_\alpha^a A_\beta = \delta_{\alpha\beta} \eta^a + 
\epsilon_{\alpha\beta\gamma} \bar{\psi}^{\gamma a},
\nonumber
\\
&\bar{Q}_\alpha^a \phi^{bc} = - \hbox{\large$\frac{1}{2}$} 
\epsilon^{ab} \psi_\alpha^c - \hbox{\large$\frac{1}{2}$} 
\epsilon^{ac} \psi_\alpha^b,
\nonumber
\\
&\bar{Q}_\alpha^a \eta^b = D_\alpha \phi^{ab} - 
\epsilon^{ab} [ V_\alpha, \rho ] + \epsilon^{ab} B_\alpha,
\nonumber
\\
&\bar{Q}_\alpha^a \psi_\beta^b = - \epsilon^{ab} F_{\alpha\beta} +
\epsilon^{ab} \epsilon_{\alpha\beta\gamma} \bar{B}^\gamma +
\delta_{\alpha\beta} \epsilon_{cd} [ \phi^{ac}, \phi^{bd} ] +
\epsilon^{ab} \epsilon_{\alpha\beta\gamma} D^\gamma \rho - 
\epsilon_{\alpha\beta\gamma} [ V^\gamma, \phi^{ab} ],
\nonumber
\\
&\bar{Q}_\alpha^a B_\beta = D_\alpha \psi_\beta^a - 
\hbox{\large$\frac{1}{2}$} D_\beta \psi_\alpha^a +
\hbox{\large$\frac{1}{2}$} [ V_\beta, \bar{\psi}_\alpha^a ] -
\epsilon_{cd} [ \phi^{ac}, \delta_{\alpha\beta} \eta^d - 
\epsilon_{\alpha\beta\gamma} \bar{\psi}^{\gamma d} ]
\nonumber
\\
&\qquad\qquad + [ \rho, \delta_{\alpha\beta} \bar{\eta}^a -
\epsilon_{\alpha\beta\gamma} \psi^{\gamma a} ] + 
\epsilon_{\alpha\beta\gamma} [ V^\gamma, \eta^a ], 
\nonumber
\\
&\bar{Q}_\alpha^a V_\beta = - \delta_{\alpha\beta} \bar{\eta}^a +
\epsilon_{\alpha\beta\gamma} \psi^{\gamma a},
\nonumber
\\
&\bar{Q}_\alpha^a \rho = \hbox{\large$\frac{1}{2}$} \bar{\psi}_\alpha^a,
\nonumber
\\
&\bar{Q}_\alpha^a \bar{\eta}^b = - \epsilon^{ab} D_\alpha \rho - 
[ V_\alpha, \phi^{ab} ] + \epsilon^{ab} \bar{B}_\alpha,
\nonumber
\\
&\bar{Q}_\alpha^a \bar{\psi}_\beta^b = - \epsilon^{ab} D_\alpha V_\beta +
\epsilon^{ab} \epsilon_{\alpha\beta\gamma} B^\gamma +
2 \delta_{\alpha\beta} [ \rho, \phi^{ab} ] - 
\epsilon_{\alpha\beta\gamma} D^\gamma \phi^{ab} -
\epsilon^{ab} \epsilon_{\alpha\beta\gamma} [ V^\gamma, \rho ] - 
\epsilon^{ab} \delta_{\alpha\beta} Y,
\nonumber
\\
&\bar{Q}_\alpha^a \bar{B}_\beta = D_\alpha \bar{\psi}_\beta^a -
\hbox{\large$\frac{1}{2}$} D_\beta \bar{\psi}_\alpha^a -
\hbox{\large$\frac{1}{2}$} [ V_\beta, \psi_\alpha^a ] -
\epsilon_{cd} [ \phi^{ac}, \delta_{\alpha\beta} \bar{\eta}^d - 
\epsilon_{\alpha\beta\gamma} \psi^{\gamma d} ]
\nonumber
\\
&\qquad\qquad - [ \rho, \delta_{\alpha\beta} \eta^a - 
\epsilon_{\alpha\beta\gamma} \bar{\psi}^{\gamma a} ] +
\epsilon_{\alpha\beta\gamma} D^\gamma \eta^a, 
\nonumber
\\
&\bar{Q}_\alpha^a Y = D_\alpha \bar{\eta}^a + [ V_\alpha, \eta^a ] + 
\epsilon_{cd} [ \phi^{ac}, \bar{\psi}_\alpha^d ] - [ \rho, \psi_\alpha^a ], 
\end{align} 
respectively. The transformation laws generated by the conjugate supercharges,
$\bar{Q}^a$ and $Q_\alpha^a$, are obtained from (\ref{4.23}) and
(\ref{4.24}) by carrying out the replacements (\ref{4.22}), mapping $Q^a$ to 
$\bar{Q}^a$ and $\bar{Q}_\alpha^a$ to $Q_\alpha^a$, respectively.

To make contact with the action and the transformations given in 
(\ref{3.10}), (\ref{3.12}) and (\ref{3.14}) we express 
(\ref{4.21}), (\ref{4.23}) and (\ref{4.24}) in terms of the complexified 
fields $A_\alpha \pm i V_\alpha$, $B_\alpha \pm i \bar{B}_\alpha$,
$\psi_\alpha^a \pm i \bar{\psi}_\alpha^a$, $\eta^a \pm i \bar{\eta}^a$
and in terms of the complexified supercharges
$Q^a \pm i \bar{Q}^a$, $\bar{Q}_\alpha^a \pm i Q_\alpha^a$, respectively. 
In addition, we combine $\phi^{ab}$ and $\rho$ to form the complex scalar 
fields 
\begin{equation*}
\zeta^{ab} = \phi^{ab} + i \epsilon^{ab} \rho
\qquad
\hbox{and}
\qquad
\bar{\zeta}^{ab} = \phi^{ab} - i \epsilon^{ab} \rho,
\end{equation*}
recalling that $\phi^{ab}$ is symmetric, $\phi^{ab} = \phi^{ba}$. Then, after 
carrying out the redefinitions
\begin{alignat*}{4}
&&\qquad
B_\alpha + i \bar{B}_\alpha &\rightarrow B_\alpha,
&\qquad
B_\alpha - i \bar{B}_\alpha &\rightarrow \bar{B}_\alpha,
&\qquad&
\\
\eta^a + i \bar{\eta}^a &\rightarrow \eta^a,
&\qquad
\eta^a - i \bar{\eta}^a &\rightarrow \bar{\eta}^a,
&\qquad
\psi_\alpha^a + i \bar{\psi}_\alpha^a &\rightarrow \psi_\alpha^a,
&\qquad
\psi_\alpha^a - i \bar{\psi}_\alpha^a &\rightarrow \bar{\psi}_\alpha^a
\\
\intertext{and}
Q^a + i \bar{Q}^a &\rightarrow Q^a,
&\qquad  
Q^a - i \bar{Q}^a &\rightarrow \bar{Q}^a,
&\qquad
\bar{Q}_\alpha^a + i Q_\alpha^a &\rightarrow \bar{Q}_\alpha^a,
&\qquad  
\bar{Q}_\alpha^a - i Q_\alpha^a &\rightarrow Q_\alpha^a,
\end{alignat*}
it is easily seen that the resulting action and transformations are precisely
the ones given in Section 4. Hence, it is proven that by a dimensional 
reduction of either of the two $N_T = 2$, $D = 4$ theories one recovers 
the $N_T = 4$ equivariant extension of the Blau--Thompson model proposed 
in Section 4. 
\bigskip
\begin{flushleft}
{\large{\bf 5. Dimensional reduction of half--twisted $N_T = 1$, 
$D = 4$ Yamron model}}
\end{flushleft}
\bigskip
After having described the $N_T = 4$ topological twist of $N = 8$, 
$D = 3$ SYM, arising from the dimensional reduction of either of the two 
$N_T = 2$, $D = 4$ theories, now, for the sake of completeness, we will also 
explicitely construct the other $N_T = 2$ topological model arising either by 
partially twisting $N = 8$, $D = 3$ SYM or by dimensional reduction of the 
half--twisted $N_T = 1$, $D = 4$ theory \cite{22} (see Diagram). 

The global symmetry group of this theory is 
$SU(2)_N \otimes \overline{SU(2)}_R$.
The action of this partially twisted theory can be described as the coupling
of the $N_T = 2$, $D = 3$ super--BF theory to a spinorial hypermultiplet
$\{ \lambda_A^{ab}, \zeta_A^a \}$. This hypermultiplet is built up from a 
$SU(2)_N \otimes \overline{SU(2)}_R$ quartet of 
Grassmann--odd spinor fields $\lambda_A^{ab}$ and a 
$\overline{SU(2)}_R$ doublet of Grassmann--even spinor fields $\zeta_A^a$. 
In order to close the topological superalgebra (see Eq. (\ref{5.30}) below), 
we introduce a $\overline{SU(2)}_R$ doublet of bosonic auxiliary spinor 
fields $Y_A^a$. The spinor indices are denoted by 
$A = 1,2$. All the spinor fields are taken in the adjoint representation of 
the gauge group $G$.

Omitting any details, after dimensional reduction of the half--twisted
$N_T = 1$, $D = 4$ theory the reduced action, with an underlying $N_T = 2$ 
off--shell equivariantly nilpotent topological shift symmetry $Q^a$, splits 
up in two $SU(2)_N \otimes \overline{SU(2)}_R$ invariant parts
\begin{equation}
\label{5.25}
S^{(N_T=2)} = S_{\rm BF} + S_{\rm M},
\end{equation}
where the first part is just the action of the $N_T = 2$, $D = 3$ super--BF 
theory given in (\ref{2.1}) and the second one is the matter action  
\begin{align}
\label{5.26}
S_{\rm M} = \int d^3x\, {\rm tr} \Bigr\{&
- i \lambda_{A ab} (\sigma^\alpha)^{AB} D_\alpha \lambda_B^{ab} - 
2 i \lambda_{A ab} (\sigma^\alpha)^{AB} [ \psi_\alpha^a, \zeta_B^b ]
\nonumber
\\
& + 2 \lambda_{A ab} [ \eta^a, \zeta^{A b} ] +
2 \epsilon_{cd} \lambda_A^{ac} [ \phi_{ab}, \lambda^{A bd} ] + 
[ \phi_{ab}, \zeta_{Ac} ] [ \phi^{ab}, \zeta^{Ac} ]
\phantom{\frac{1}{2}}
\nonumber
\\
& - i \epsilon_{ab} \zeta_A^a (\sigma^\alpha)^{AB} [ B_\alpha, \zeta_B^b ] - 
2 i \epsilon_{ab} Y_A^a (\sigma^\alpha)^{AB} D_\alpha \zeta_B^b + 
2 Y^{A a} Y_{A a} \Bigr\};
\end{align}
here, $D_\alpha$ is the covariant derivative (in the adjoint representation)
and $\sigma_\alpha$ are the Pauli matrices,
\begin{equation*}
(\sigma_\alpha)_A^{~~\!C} (\sigma_\beta)_{CB} = 
\delta_{\alpha\beta} \epsilon_{AB} +
i \epsilon_{\alpha\beta\gamma} (\sigma^\gamma)_{AB},
\qquad
\alpha = 1,2,3.
\end{equation*}
The spinor index $A$ is raised and lowered as follows, 
$\varphi_A = \varphi^B \epsilon_{BA}$ and 
$\varphi^A = \epsilon^{AB} \varphi_B$ with 
$\epsilon^{AC} \epsilon_{CB} = - \delta^A_B$. 

The action (\ref{5.26}) can be cast in the $Q^a$--exact form
\begin{equation*}
S_{\rm M} = \hbox{\large$\frac{1}{2}$} \epsilon_{ab} Q^a Q^b X_{\rm M}
\end{equation*}
with the gauge boson
\begin{equation*}
X_{\rm M} = - \hbox{\large$\frac{1}{2}$} \int d^3x\, {\rm tr} \Bigr\{
\lambda_{A ab} \lambda^{A ab} + 
i \epsilon_{ab} \zeta_A^a (\sigma^\alpha)^{AB} D_\alpha \zeta_B^b \Bigr\}.
\end{equation*}

By a straightforward calculation in can be verified that both 
parts in (\ref{5.25}) are separately invariant under the off--shell 
equivariantly nilpotent topological supersymmetry $Q^a$ 
(cf.,~Eqs.~(\ref{2.2})),
\begin{alignat}{2}
\label{5.27}
&Q^a A_\alpha = \psi_\alpha^a,
&\qquad
&Q^a \zeta_A^b = \lambda_A^{ab},
\nonumber
\\
&Q^a \phi^{bc} = \hbox{\large$\frac{1}{2}$} \epsilon^{ab} \eta^c +
\hbox{\large$\frac{1}{2}$} \epsilon^{ac} \eta^b,
&\qquad
&Q^a \eta^b = - \epsilon_{cd} [ \phi^{ac}, \phi^{bd} ],
\nonumber
\\
&Q^a \psi_\alpha^b = D_\alpha \phi^{ab} + \epsilon^{ab} B_\alpha,
&\qquad
&Q^a B_\alpha = - \hbox{\large$\frac{1}{2}$} D_\alpha \eta^a -
\epsilon_{cd} [ \phi^{ac}, \psi_\alpha^d ],
\nonumber
\\
&Q^a \lambda_A^{bc} = - [ \phi^{ab}, \zeta_A^c ] + \epsilon^{ab} Y_A^c,
&\qquad
&Q^a Y_A^b = \hbox{\large$\frac{1}{2}$} [ \eta^a, \zeta_A^b ] -
\epsilon_{cd} [ \phi^{ac}, \lambda_A^{db} ],
\end{alignat}
and under the vector supersymmetry $\bar{Q}_\alpha^a$ 
(cf.,~Eqs.~(\ref{2.3})),
\begin{align}
\label{5.28}
&\bar{Q}_\alpha^a A_\beta = \delta_{\alpha\beta} \eta^a + 
\epsilon_{\alpha\beta\gamma} \psi^{\gamma a},
\nonumber
\\
&\bar{Q}_\alpha^a \zeta_A^b = i (\sigma_\alpha)_{AB} \lambda^{B ab},
\nonumber
\\
&\bar{Q}_\alpha^a \phi^{bc} = - \hbox{\large$\frac{1}{2}$} 
\epsilon^{ab} \psi_\alpha^c - \hbox{\large$\frac{1}{2}$} 
\epsilon^{ac} \psi_\alpha^b,
\nonumber
\\
&\bar{Q}_\alpha^a \eta^b = D_\alpha \phi^{ab} + \epsilon^{ab} B_\alpha,
\nonumber
\\ 
&\bar{Q}_\alpha^a \psi_\beta^b = - \epsilon^{ab} F_{\alpha\beta} +
\epsilon^{ab} \epsilon_{\alpha\beta\gamma} B^\gamma - 
\epsilon_{\alpha\beta\gamma} D^\gamma \phi^{ab} +
\delta_{\mu\nu} \epsilon_{cd} [ \phi^{ac}, \phi^{bd} ],
\nonumber
\\
&\bar{Q}_\alpha^a B_\beta = D_\alpha \psi_\beta^a - 
\hbox{\large$\frac{1}{2}$} D_\beta \psi_\alpha^a +
\epsilon_{\alpha\beta\gamma} D^\gamma \eta^a +
\epsilon_{cd} [ \phi^{ac}, \epsilon_{\alpha\beta\gamma} \psi^{\gamma d} - 
\delta_{\alpha\beta} \eta^d ], 
\nonumber
\\
&\bar{Q}_\alpha^a \lambda_A^{bc} = - \epsilon^{ab} D_\alpha \zeta_A^c +
i (\sigma_\alpha)_{AB} ( [ \phi^{ab}, \zeta^{B c} ] + \epsilon^{ab} Y^{B c} ),
\nonumber
\\
&\bar{Q}_\alpha^a Y_A^b = D_\alpha \lambda_A^{ab} + 
\hbox{\large$\frac{1}{2}$} [ \psi_\alpha^a, \zeta_A^b ] -
i (\sigma_\mu)_{AB} ( [ \eta^a, \zeta^{B b} ] -
\epsilon_{cd} [ \phi^{ac}, \lambda^{B db} ] ).
\\
\intertext{Furthermore, it can be proven that the sum of the two parts 
in (\ref{5.25}) is invariant under the transformations generated by the 
spinorial supercharges $Q_A^{ab}$,}
\label{5.29}
&Q_A^{ab} A_\alpha = i (\sigma_\alpha)_{AB} \lambda^{B ab},
\nonumber
\\
&Q_A^{ab} \zeta_B^c = \epsilon_{AB} \epsilon^{bc} \eta^a -
i (\sigma^\alpha)_{AB} \epsilon^{bc} \psi_\alpha^a, 
\nonumber
\\
&Q_A^{ab} \phi^{cd} = \hbox{\large$\frac{1}{2}$}
\epsilon^{ac} \lambda_A^{db} + \hbox{\large$\frac{1}{2}$}
\epsilon^{ad} \lambda_A^{cb},
\nonumber
\\
&Q_A^{ab} \eta^c = [ \phi^{ac}, \zeta_A^b ] - \epsilon^{ac} Y_A^b,
\nonumber
\\
&Q_A^{ab} \psi_\alpha^c = - \epsilon^{ac} D_\alpha \zeta_A^b +
i (\sigma_\alpha)_{AB} ( [ \phi^{ac}, \zeta^{B b} ] + \epsilon^{ac} Y^{B b} ),
\nonumber
\\
&Q_A^{ab} B_\alpha = \hbox{\large$\frac{1}{2}$} D_\alpha \lambda_A^{ab} + 
[ \psi_\alpha^a, \zeta_A^b ] -
i (\sigma_\alpha)_{AB} ( [ \eta^a, \zeta^{B b} ] -
\epsilon_{cd} [ \phi^{ac}, \lambda^{B db} ] ),
\nonumber
\\
&Q_A^{ab} \lambda_B^{cd} = \epsilon^{ac} [ \zeta_A^b, \zeta_B^d ] +
\epsilon_{AB} [ \phi^{ac}, \phi^{bd} ] +
i (\sigma^\alpha)_{AB} \epsilon^{bd} ( D_\alpha \phi^{ac} - 
\epsilon^{ac} B_\alpha ),
\nonumber
\\
&Q_A^{ab} Y_B^c = - \hbox{\large$\frac{1}{2}$} 
[ \lambda_A^{ab}, \zeta_B^c ] +
[ \lambda_B^{ac}, \zeta_A^b ] - 
\epsilon_{AB} [ \eta^a, \phi^{bc} ] -
i (\sigma^\alpha)_{AB} \epsilon^{bc} (
D_\alpha \eta^a + \epsilon_{ef} [ \phi^{ae}, \psi_\alpha^f ] ).
\end{align}
By the symmetry requirements 
$Q^a S_{\rm BF} = \bar{Q}_\alpha^a S_{\rm BF} = 0$ and
$Q^a S_{\rm M} = \bar{Q}_\alpha^a S_{\rm M} = 0$, together with
$Q_A^{ab} ( S_{\rm BF} + S_{\rm M } ) = 0$, all the relative
numerical coefficients of the action $S_{\rm BF} + S_{\rm M}$ are 
uniquely fixed, execpt for a single overall coupling constant. 

The eight supercharges $Q^a$, $\bar{Q}_\alpha^a$ and $Q_A^{ab}$, 
together with the generator $P_\alpha$ of space--time translations, satisfy 
the topological superalgebra
\begin{align}
\label{5.30}
\{ Q^a, Q^b \} &= - 2 \delta_G(\phi^{ab}),
\nonumber
\\
\{ Q^a, \bar{Q}_\alpha^b \} &= \epsilon^{ab} (
- i P_\alpha + \delta_G(A_\alpha) ),
\nonumber
\\
\{ Q^a, Q_A^{bc} \} &= - \epsilon^{ab} 
\delta_G(\zeta_A^c),
\nonumber
\\
\{ \bar{Q}_\alpha^a, \bar{Q}_\beta^b \} &\doteq - 2 \delta_{\alpha\beta} 
\delta_G(\phi^{ab}),
\nonumber
\\
\{ \bar{Q}_\alpha^a, Q_A^{bc} \} &\doteq - i \epsilon^{ab} 
(\sigma_\alpha)_{AB} \delta_G(\zeta^{B c}),
\nonumber
\\
\{ Q_A^{ac}, Q_B^{bd} \} &\doteq 2 \epsilon^{cd} \epsilon_{AB}
\delta_G(\phi^{ab}) - i \epsilon^{ab} \epsilon^{cd} (\sigma^\alpha)_{AB} (
- i P_\alpha + \delta_G(A_\alpha) ).
\end{align}
Finally, let us notice that the dimensional reduction of the 
half--twisted $N_T = 1$, $D = 4$ theory can also be described by  
decomposing $Spin(7) \rightarrow G_2 \rightarrow 
SU(2)_N \otimes \overline{SU(2)}_R$ \cite{16}. In that description, however,
only the diagonal subgroup of the global symmetry group
$SU(2)_N \otimes\break \overline{SU(2)}_R$ is manifest. The action of this 
theory coincides with the dimensional reduction of the $N_T = 1$, $D = 4$ 
Donaldson--Witten theory \cite{3} coupled to the standard hypermultiplet 
\cite{32} to $D = 3$.
\bigskip
\begin{flushleft}
{\large{\bf Conclusions and Remarks}}
\end{flushleft}

In the present paper we proposed a $N_T = 4$ equivariant extension of the
Blau--Thompson $N_T = 2$ non--equivariant topological model in $D = 3$ 
Euclidean space--time. Furthermore, we showed, by proving the equivalence 
with the dimensional reduction to $D = 3$ of the Yamron--Vafa--Witten 
$N_T = 2$, $D = 4$ theory, that this extended topological model coincides 
with one of the two inequivalent topological models which, according to the
classification of \cite{16}, may be constructed by twisting the $N = 8$, $D = 3$ 
super--Yang--Mills theory. In addition, we constructed explicitly also the 
other topological model which may by obtained, namely, the $N_T = 2$, $D = 3$ 
super--BF theory coupled to a spinorial hypermultiplet. 

All these models, including the $N_T = 2$, $D = 3$ super--BF model,
are considered in the flat Euclidean space--time where, besides the 
topological shift symmetry also the vector supersymmetry can be constructed. 
All these symmetries have been given explicitely. The actions of the 
corresponding theories are shown to be uniquely determined, up to an overall 
coupling constant, when they are required to be invariant not only with 
respect to the topological shift symmetry but also with respect to the vector 
supersymmetry. Furthermore, it is shown that these symmetry operations, 
together with the generators of space--time translations, fulfill 
corresponding topological superalgebras.

As already mentioned in the introduction, the Blau--Thompson $N_T = 2$ 
non--equivariant topological model and its $N_T = 4$ equivariant
extension provide not only the link between the various topological theories 
arising from twisting $N = 1$, $D = 6$ or $N = 1$, $D = 10$ super--Yang--Mills 
theory, but also are pre--candidates, after carrying out a dimensional reduction 
to $D = 2$, for Hodge--type cohomological theories in $D = 2$. Of course,
in searching for all of the Hodge--type cohomological theories in $D = 2$
a complete group theoretical classification of the topological models in
$D = 2$ and their explicit construction would be necessary.



\begin{thebibliography}{99}
\small{
\bibitem{1} A. S. Schwarz,
            {\it Lett. Math. Phys.} {\bf 2} (1978) 247
\bibitem{2} E. Witten,
            {\it J. Diff. Geom.} {\bf 17} (1982) 661;
            {\it Nucl. Phys.} {\bf B 202} (1982) 253
\bibitem{3} E. Witten,
            {\it Commun. Math. Phys.} {\bf 117} (1988) 353
\bibitem{4} E. Witten,
            {\it Commun. Math. Phys.} {\bf B 121} (1989) 351
\bibitem{5} J. Distler,
            {\it Nucl. Phys.} {\bf B 342} (1990) 523
\bibitem{6} E. Witten,
            {\it Phys. Lett.} {\bf 206} (1988) 601
\bibitem{7} E. Witten,
            {\it Nucl. Phys.} {\bf B 311} (1988) 46;
            {\it Nucl. Phys.} {\bf B 340} (1990) 281
\bibitem{8} J. Labastida, J. Pernici and E. Witten,
            {\it Nucl. Phys.} {\bf B 310} (1988) 611
\bibitem{9} S. Donaldson,
            {\it J. Diff. Geom.} {\bf 18} (1983) 279;
            {\it Topology.} {\bf 29} (1990) 257
\bibitem{10} V. Jones,
            {\it Bull. Am. Math. Soc.} {\bf 12} (1985) 103
\bibitem{11} L. Baulieu and I. M. Singer,
             {\it Nucl. Phys.} (Proc. Suppl.) {\bf B 5} (1988) 12
\bibitem{12} A. Floer,
             {\it J. Diff. Geom.} {\bf 30} (1989) 207;
             {\it Commun. Math. Phys.} {\bf 118} (1988) 215
\bibitem{13} For review, see, e.g., 
             D. Birmingham, M. Blau, M. Rakowski and G. Thompson,
             {\it Physics Reports} {\bf 209} (1991) 129
\bibitem{14} E. Witten,
             {\it Commun. Math. Phys.} {\bf B 118} (1988) 411
\bibitem{15} M. Bershadsky, V. Sadov and C. Vafa,
             {\it Nucl. Phys.} {\bf B 463} (1996) 420
\bibitem{16} M. Blau and G. Thompson, 
             {\it Nucl. Phys.} {\bf B 492} (1997) 545,
	     {\it Phys. Lett.} {\bf B 415} (1997) 242
\bibitem{17} J. A. Minahan, D. Nemeschansky, C. Vafa and N. P. Warner,
             {\it Nucl. Phys.} {\bf B 527} (1998) 581
\bibitem{18} C. Vafa and E. Witten,
             {\it Nucl. Phys.} {\bf B 431} (1994) 3
\bibitem{19} M. Bershadsky, A. Johansen, V. Sadov and C.Vafa, 
             {\it Nucl. Phys.} {\bf B 448} (1995) 116  
\bibitem{20} J. Maldacena,
             {\it Adv. Theor. Math. Phys.} {\bf 2} (1998) 231
\bibitem{21} B. Geyer and D. M\"ulsch, 
             {\it Twisted $N = 8$, $D = 2$ super--Yang-Mills theory
             as example of a Hodge--type cohomological theory},
	     in preparation
\bibitem{22} J. Yamron,
             {\it Phys. Lett.} {\bf B 213} (1988) 325
\bibitem{23} N. Marcus,
             {\it Nucl. Phys.} {\bf B 452} (1995) 331
\bibitem{24} E. Witten,
             {\it Nucl. Phys.} {\bf B 323} (1989) 113
\bibitem{25} D. Birmingham, M. Blau and G. Thompson,
             {\it Int. J. Mod. Phys.} {\bf A 5} (1990) 4721
\bibitem{26} M. Blau and G. Thompson,
             {\it Commun. Math. Phys.} {\bf 152} (1993) 41                 
\bibitem{27} N. Seiberg and E. Witten,
             {\it Gauge Dynamics and Compactifications to Three Dimensions},
             in: {\sf The mathematical beauty of physics},
	     Saclay 1996, hep-th/9611230
\bibitem{28} J.C. Wallet,
             {\it Phys. Lett.} {\bf B 235} (1990) 71
\bibitem{29} D. Birmingham, M. Blau and G. Thompson,
             {\it Int. J. Mod. Phys.} {\bf A 5} (1990) 4721
\bibitem{30} R. Dijkgraaf and G. Moore,
             {\it Commun. Math. Phys.} {\bf 185} (1997) 411
\bibitem{31} C. Lozano,
             {\it Duality in Topological Quantum Field Theories},
             PhD thesis, 1999, hep-th/9907123
\bibitem{32} P. Fayet,
             {\it Nucl. Phys.} {\bf B113} (1976) 135
}
\end{thebibliography}
\end{document}